\documentclass{article}
\usepackage[margin=25mm]{geometry}
\usepackage{graphicx} 
\usepackage{caption}
\usepackage{amsmath}
\usepackage{amssymb}
\usepackage[affil-it]{authblk}
\title{Population and Inequality Dynamics in Simple Economies}
\author[1,*]{John Stevenson}
\date{}
\affil[1]{Independent}
\affil[*]{Corresponding Author: John Stevenson, jcs@alumni.caltech.edu}
\begin{document}

\maketitle

\begin{abstract}
While the use of spatial agent-based and individual-based models has flourished across many scientific disciplines, the complexities these models generate are often difficult to manage and quantify. This research reduces population-driven, spatial modeling of individuals to the simplest configurations and parameters: an equal resource opportunity landscape with equally capable individuals; and asks the question, "Will valid complex population and inequality dynamics emerge from this simple economic model?"  Two foraging economies are modeled: subsistence and surplus. The resulting, emergent population dynamics are characterized by their sensitivities to agent and landscape parameters. The various steady and oscillating regimes of single-species population dynamics are generated by appropriate selection of model growth parameters. These emergent dynamics are shown to be consistent with the equation-based, continuum modeling of single-species populations in biology and ecology. The intrinsic growth rates, carry capacities, and delay parameters of these models are implied for these simple economies. Aggregate measures of individual distributions are used to understand the sensitivities to model parameters. New local measures are defined to describe complex behaviors driven by spatial effects, especially extinctions. Inequality measures are defined and applied to surplus economies. Sensitivities of inequality dynamics to model parameters are explored. Individual wealth trajectories are examined for insight into the different dynamics of inequality distributions. This simple economic model is shown to generate significantly complex population and inequality dynamics. Model parameters generating the intrinsic growth rate have strong effects on these dynamics, including large variations in inequality. Significant inequality effects are shown to be  mitigated by birth costs above and beyond their contribution to the intrinsic growth rate. The highest levels of inequality are found during the initial non-equilibrium period and are driven by factors different than those driving steady state inequality.
\end{abstract}

\section{Introduction}

\begin{quote}
\itshape
"It can scarcely be denied that the supreme goal of all theory is make the irreducible basic elements as simple and as few as possible without having to surrender the adequate representation of a single datum of experience."

Albert Einstein (Caprice, 2011)
\end{quote}

The modeling of populations of individuals over time on a spatial grid has become a powerful tool in both the life sciences (DeAngelis and Grimm, 2014; Badham et. al., 2018; Shoukat and Moghadas, 2020), and the social sciences (Heppenstall, 2020; Chen, 2011). This modelling began in the field of ecology, where it is called Individual-Based Modelling, and then expanded into the social sciences, where it is called Agent-Based Modelling. Many of these models, particularly in biology and ecology, are population driven, which exerts selection pressure on the characteristics of populations (Gause, 1934). The ability of these models to support increasingly complicated behaviors (both programmed and learned), environments, and sensory inputs has resulted in investigations that are both broad, across many scientific disciplines (Vincenot 2018), and detailed (di Porcia e Brugnera, 2020; Kohler and Gumerman, 2000). The field has come to recognize a variety of problems that have accompanied the growth of this new computational approach: reproducibility, validation and verification, sensitivity analysis, and communication of results, to name just a few. (Mason, 2020; Lee et. al, 2015; Angus and Hassani-Mahmooei, 2015). By using a simple economic model, this research provides an opportunity to clearly address these issues in a less complicated environment. By understanding and communicating the simulation process, the dynamics and sensitivities of the model to its parameters, and the resulting unexpected insights into the transient characteristics of non-equilibrium population growth, a foundation is built to support research of more complex, population-driven models. 

These efforts start by simplifying the first and simplest population-driven model found in Epstein and Axtell's classic romp through agent-based modelling for social science (Epstein and Axtell, 1996). Their simple model, with two rules for agents and one rule for the landscape, consists of a heterogeneous population of agents who can look, move, eat, harvest, die and reproduce on a two-dimensional grid upon which food grows. The agents' abilities are  heterogeneous, differing in metabolism, initial endowment, vision, movement and fertility. The landscape grows food at a fixed rate throughout the field but differs in the capacity of a given location to hold food. This difference in spatial capacity leads to "deserts" and "mountains" of food availability. 

This research makes significant simplifications to their already simple model. The properties of agents are taken to be homogeneous, each agent having identical properties. Reproduction is simplified to be asexual. The maximum food capacity of the landscape is identical throughout the field. There are no endowments granted to agents at any point. These simplifications create a model that has equal opportunity for all its equally capable agents, removing some confounding factors in understanding wealth inequality. While these simplifications preserve the relevant emergent behaviors of Epstein and Axtell's model: oscillating and chaotic regimes, extinctions, and perhaps most surprisingly, significantly unequal wealth distributions; they also allow investigations into transient dynamics of population characteristics, validation and verification of the simulation through both writing the conservation of energy equations and mapping of the intrinsic growth rates and dynamic regimes to the established models of biology and ecology, analysis of sensitivities over a manageable set of model parameters, and, finally, an appreciation of the emergent complexity of even this simplified economic model. And this approach allows posing of the basic question: will complex inequalities arise in populations of identical agents presented with equal opportunities?

This issue of inequality among equals arose in discussions of fairness of tax codes (Musgrave 1959, Feldstein 1976). Horizontal equity was defined as an objective for taxes such that those taxes, when applied to individuals equally well off, left those individuals equally well off. Furthermore, the newly applied tax should not change the "utility" ranking of the individuals (Rosen 1978). The term was refined as horizontal inequality to measure change in household welfare rather than just income ranking (Slesnick 1989). Foley (1999) identified horizontal inequalities for equal traders in his statistical equilibrium analyses of a commodity market. This research presents evidence for horizontal inequality among equal agents with equal opportunity using a simple economic model under both equilibrium and non-equilibrium conditions.

The computational process and parameters for this simple economic model are provided in sufficient detail to ensure reproducibility [Section 2.1]. This computational model is called a simple economic model whereas a specific simulation is referred to as a simple economy [Table 2]. The equations for conservation of resources in the landscape and in the population are written to provide concise and accurate descriptions of the computational algorithms, and to ensure the computational process properly accounts for the conservation of energy (resources) [Section 2.2].  Examination of the sensitivities of these simple economies to their parameters produces various dynamic regimes of population trajectories [Section 2.4]. The credibility of these population-driven dynamics is supported by comparisons with the extensive literature of mathematical biology and ecology on equation-based models for single-species populations. Implied intrinsic growth rates, carry capacities, and delay parameters are computed using these equation-based models [Section 2.5]. 

One of the advantages of population models based on individuals is the ability to examine both global, aggregate behavior as well as local behaviors. These advantages are used to explore the local and global ways in which populations become extinct [Section 3]. Furthermore, calculation of the distribution of populations' surplus resources allows an in depth measurement of varying degrees of inequality generated by different simple economies. The factors driving these inequalities are identified and the non-equilibrium state drivers are shown to be quite different from the steady state drivers [Section 4].  

Simple economies with equal opportunity environments and equally capable individuals generate complex wealth distributions whose inequalities are dependent on the intrinsic growth rate of the population, the cost of reproduction, and whether the economy has reached equilibrium. 

\section{Model of Simple Economies}

The computational model is described in sufficient detail to ensure reproducibility. Equations for the conservation of energy (resources) are developed to both precisely define the computational process and to validate the conservation of resources. Subsistence and surplus economies are defined, actual and theoretical carry capacities are calculated, and sensitivities of actual carry capacities to model parameters are developed. The model parameters are further simplified and sensitivities of population dynamics to these remaining parameters are explored. The population dynamics are put in context with the single-species, equation-based models of mathematical biology and the intrinsic growth rates, carry capacities, and delay parameters for simple economies are estimated.

\subsection{Computational Model and Process}

Table 1 provides the definition of the agents' and landscape's parameters for the simple economic model. Vision and movement are along rows and columns only. The two dimensional landscape wraps around the edges (often likened to a torus). Agents are selected for action in random order each cycle. The selected agent moves to the closest visible cell with the most resources with ties resolved randomly. After movement, the agent harvests and consumes (metabolizes) the required resources. At this point, if the agent's resources are depleted, the agent is removed from the landscape. Otherwise an agent of sufficient age (puberty) then considers reproduction, requiring sufficient resources (birth cost), a lucky roll of the fertility die (infertility), and an empty von Neumann neighbor cell, which are only the four neighboring spaces one step away by row or column. If a birth occurs in a configuration with zero puberty, the newborn is added to the list of agents to be processed in this current cycle. Otherwise (puberty $>0$), the newborn is placed in the empty cell and remains inert until the next action cycle. With this approach for the action cycle, no endowments are required whether for new births or for the agent(s) at start-up. Once all the agents have cycled through, the landscape replenishes at the growth rate and the cycle ends.

\begin{table}[h!]
\centering
 \begin{tabular}{|c|c|c|c|c|}
   \hline
    Agent Characteristic & Notation & Range & Units & Purpose \\
   \hline
   vision & $v$ &  1-25 &  cells & vision of resources on landscape \\
    movement & -- &  1-25 &  cells per cycle &  movement about landscape \\
metabolism & $m$ & 1+ & resources per cycle &  consumption of resource \\
birth cost & $bc$ & 0+ & resources &  sunk cost for reproduction \\
infertility & $f$ & 1+ & 1/probability & likelihood of birth \\
puberty & $p$ & 0+ &  cycles &  age to start reproduction\\
surplus & $S$ & 0+ & resources &  storage of resource across cycles \\
\hline
 \end{tabular}
   \bigbreak
 \begin{tabular}{|c|c|c|c|}
   \hline
Landscape Characteristic & Notation & Value & Units\\
\hline
rows & -- & 50 & cells \\
columns & -- & 50 & cells\\
max capacity &$R$ & 4 & resource per cell\\
growth & $g$ & 1 & resource per cycle per cell \\
initial & $R_{0}$ & 4 & resource, all cells\\
\hline
 \end{tabular}
 \caption{Agent and Landscape Parameters of the Simple Economic Model}
 \label{Table 1:}
\end{table}
Two metabolism rates ($m$), uniform across a given population, are modelled: one that consumes per cycle the maximum capacity of resource per cell and one that consumes per cycle 25\% less than the maximum capacity per cell. The former is considered a subsistence economy ($m=4$) and the latter a surplus economy ($m=3$). And, again for simplicity, the vision and movement characteristics are set to equal values of distance.

The initial population starts with a single agent with no initial endowment. The population trajectory over time is dependent on the three growth characteristics, infertility, puberty, and birth cost. The birth cost parameter represents the amount of resources the parent agent expends to reproduce. The surplus of the parent agent must be sufficient to provide this amount and it is a sunk cost. The puberty parameter represents the number of cycles required after birth before the new agent can reproduce. The infertility parameter ($f$) defines the probability, $P_{b}$, that an agent with sufficient resources and space (a free neighboring space) will, in fact, reproduce in this cycle as:

\begin{equation}
P_{b}=1/f
\end{equation}
where $f>=1$. This parameter can be seen as a simple way to approximate various more complex reproductive characteristics and intrinsic growth rates.

Since these parameters will be used to generate many illustrative simple economies, Table 2 provides a shorthand for labeling these parameters for the simple economic model:

\begin{table}[h!]
\centering
 \begin{tabular}{|c|c|c|}
   \hline
    Simple Economy Parameters & symbol & label \\
   \hline
infertility & $f$ & fXX \\
puberty & $p$ & pX \\
birth cost & $bc$ & bcXX \\
metabolism & $m$ & mX \\
simulation length & $T$ & \_XX \\
   \hline
 \end{tabular}
 \caption{Shorthand Labeling for Simple Economies. As an example, f10p1bc40m3\_50K is a surplus economy ( $m$ of 3 resources per cycle) with $f$ at 10, $p$ at 1 cycle, $bc$ at 40 resources per birth, and simulation length of 50,000 cycles.}
 \label{Table 2:}
\end{table}

\subsection{Conservation of Energy}
The calculation of conservation of energy (resources) confirms the validity of the simulation and provides a precise description of the computational process. Two energy conservation equations are written, one for the landscape $E_{L}$, and one for the population $E_{P}$. In order to easily write the conservation equations, the energy harvested $H(t)$ from the landscape by the agents at time $t$ is defined as:
\begin{equation}
H(t) = \sum_{a=1}^{A(t-1)}r[c(a),t]+\delta_{p_{0}}\sum_{a=1}^{B(t)}r[c(a),t]
\end{equation}
where $a$ is the agent index, $A(t-1)$ is the list of agents alive at the end of the previous cycle $t-1$, $B(t)$ is the list of new agents generated in this cycle $t$, $\delta_{p_{0}}$ is one if $p$ equals 0 and 0 otherwise, $c(a)$ is the cell location of an agent indexed as $a$, and $r[c(a),t]$ are the resources in the cell occupied by $a$ that are harvested by $a$ in the current cycle.
The landscape resource conservation equation can now be written as:
\begin{equation}
\Delta E_{L}(t)= \sum_{c=1}^{N_{c}}g_{c}(t-1)-H(t)
\end{equation}
where $\Delta E_{L}$ is the change in total resources of the landscape from the previous cycle $t-1$ to the current cycle $t$, $N_{c}$ is the number of cells in the landscape, $g_{c}(t-1)$ is the resource added to cell $c$ at the end of the previous cycle $t-1$. $g_{c}(t)$ is given as:
\begin{equation}
	g_{c}(t) =  
	\begin{cases}
		g & r[c,t]+g\le R\\
		R-r[c,t] & r[c,t]+g>R\\
	\end{cases}
\end{equation}	 
where $R$ is the maximum resources in a landscape's cell, and $g$ is the growth rate of resources in landscape's cell. The conservation equation for the resources in the population can now be written as:
\begin{equation}
	\Delta E_{P}(t) = H(t)-\sum_{a=1}^{A(t)}m -\sum_{a=1}^{D(t)}(S_{a}(t)+m)-\sum_{a=1}^{B(t)}[bc -\delta_{p_{0}}m]
\end{equation}
where $\Delta E_{P}(t)$ is the change in surplus resources stored in the population from the previous cycle to the current cycle, $A(t)$ is the list of agents alive at time $t$, $m$ is the (constant) metabolism, $bc$ is the (constant) birth cost, $D(t)$ is the list of agents that died in this cycle and $S_{a}(t)$ is the surplus resources of those agents $a$ on list $D$ which have died ($S_{a}(t)<0$) in this cycle so that $S_{a}(t)+m$ are the (positive) resources lost upon its death.

Equation 5 shows that in addition to the energy consumed by metabolism, there are two additional terms that remove energy from the system and, therefore, represent sunk costs. The first sunk cost ($\sum_{a}^{D(t)}(S_{a}(t)+m)$) represents the stored, surplus resources of those agents that died during this cycle. The second sunk cost ($\sum_{a=1}^{B(t)}[bc -(!\delta_{p})m]$) represents the cost of reproducing the new agents in this cycle. The relationship between these two terms and their relative sizes play an important role in inequality distributions [Section 4]. 

Table 3 provides example energy balances for representative simple economies. The sum of changes in resources on the landscape (Equation 3) and in the population (Equation 5) are compared to the actual values of these resources at the end of the simulations of representative simple economies. These results confirm the above equations accurately describe the computational process and verify that this process conforms to the conservation of energy (resources).

\begin{table}[h!]
\centering
\begin{tabular}{|c|c|c|c|c|}
  \hline
     &  &  & & \\
    Simple Economy & $\sum_{1}^{T}\Delta E_{L}$ & $\sum_{1}^{T}\Delta E_{P}$ & Landscape Resources at $T$ & Population Resources at $T$\\
   \hline
f85p1bc0m3\_50K & 2,603 & 4,861 & 2,603 & 4,861\\
f10p1bc0m3\_10K & 2,464 & 646 & 2,464 & 646\\
f10p1bc40m3\_50K & 2,607 & 16,608 & 2,607 & 16,608\\
f1p0bc0m4\_6 & 3,537 & 0 & 3,537 & 0 \\
   \hline
 \end{tabular}
 \caption{Examples of the Energy (Resource) Balance. The energy balances for representative simple economies. The sum of changes in resources on the landscape (Equation 3) and in the population (Equation 5) are compared to the actual values of these resources at the end of the simulations.}
 \label{Table 3:}
\end{table}

\subsection{Carry Capacity}
The carry capacity of a landscape is a parameter that self-limits the growth of a single-species population. It has both a physical interpretation through the conservation of resources and a modelling interpretation for the application of equation-based models of populations. The theoretical carry capacity is developed from the conservation equations and the actual carry capacities are computed for various parameters of the simple economic model. 

\subsubsection{Theoretical Carry Capacity}

The theoretical carry capacity ($K_{T}$) for any simple economy is based on balancing the resources generated per cycle ($\Delta E_{L}$) with the resources consumed by the population ($\Delta E_{P}$) in that cycle assuming no deaths or births. Furthermore, the growth of resources assumes that every cell in the landscape is below the maximum capacity $R$. Setting Equation 3 equal to Equation 5 with these assumptions and solving for $K_{T}$ yields:
\begin{equation}
K_{T}=\frac{N_{c}}{\bar m}*g
\end{equation}
where $N_{c}$ is the number of cells in the landscape, $g$ is resource growth per cell per cycle, and $\bar m$ is the mean metabolism of the population. The growth rate of resources is uniform across the landscape, providing equal opportunity to all locations. The metabolism, along with vision/movement, is uniform across the entire population, providing equal ability to all agents.

A simple subsistence economy has an uniform agent metabolism of four resource units per cycle for each agent, yielding a theoretical carry capacity of 625 agents. Since the maximum capacity of a landscape cell is also four resource units, the agents in this economy never have the opportunity to store resources from one cycle to the next. A simple surplus economy, on the other hand, with an uniform agent metabolism of three units per cycle for each agent, yields a theoretical carry capacity of 833 agents. This economy does provide for the storage of surplus resources from one cycle to the next and there is no limit to how many resources can be stored by an agent.

\subsubsection{Actual Carry Capacity}

The sensitivity of actual carry capacity to vision/movement with no initial stores is shown in Figure 1. The actual carry capacity is the mean of the population at steady state. The standard deviation of these means are also reported in this figure. The population mean and standard deviation are computed in the steady state window (2,000 through 3,000 cycles for the runs with infertility of 85; and 400 through 1000 cycles for the others). Determination of when steady state is achieved is discussed in detail in the following sections.

\begin{center}
	\centering
	\includegraphics[angle=-90,scale=0.8]{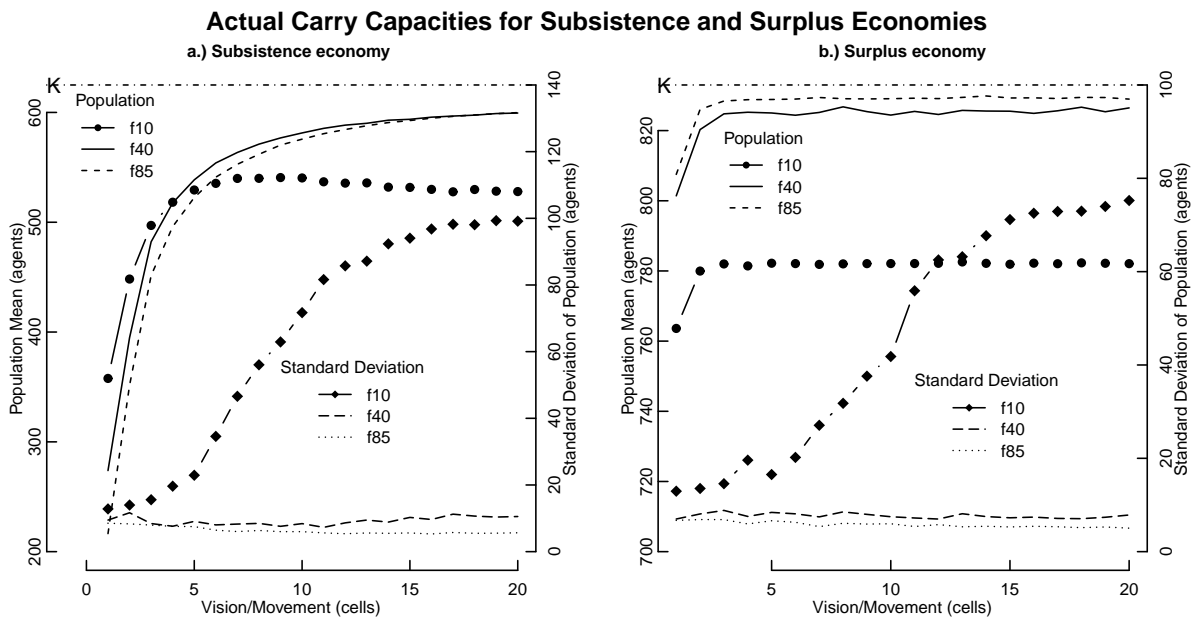}
	\captionof{figure} {a.) The means and standard deviations of population level for three subsistence (a) and surplus (b) economies by infertility with puberity ($p=1$) and birth cost ($bc=0$) held constant. $K$ represents the carry capacity of the landscape for each economy and f represents the infertility parameter. The means and standard deviations are taken over a steady state time window (2,000 through 3,000 cycles for the runs with infertility of 85; and 400 through 1000 cycles for the others). Due to the lack of any surplus resources, the subsistence economies struggle to reach the theoritical carry capacity. The f10p1bc0 parameters for both surplus and subsistence economies show the lowest actual carry capacities and an order of magnitude increase in volatility as the vision/movement increases. These increases indicate a transition to an unstable regime.} 
\end{center}
                     
The actual carry capacity falls short of the theoretical capacity of the landscape due to local, spatial effects. These local inefficiencies have greater effect on the subsistence economy due to the agents lack of surplus resources to carry-over from one cycle to the next. All subsistence economies, regardless of infertility, have trouble approaching the theoretical carry capacity of the landscape with reasonable vision and movement. For both economies, the standard deviations of the means of the populations with infertility 10 (f10) spike up above the standard deviations of populations with higher infertility. This transition to an order of magnitude higher variance is characteristic of a change in the dynamic regime of the economy defined by these model parameters. These regime changes play an important role throughout this research.

Based on the relatively flat sensitivity of the carry capacity to vision/movement once the agents can see and move more than five cells, the vision and movement parameters will be held constant at $v=6$ for the simple economies of the remaining experiments. Model parameters that remain variable are part of either the reproductive process ($f, p, bc$) or define the simple economic model as subsistence or surplus based on metabolism ($m$). 

\subsection{Sensitivity of Population Dynamics to Model Properties}

Essential to understanding population dynamics generated by this simple economic model are the relationships of emergent dynamics to the remaining model parameters. The sensitivities of population dynamics of both surplus and subsistence economies are computed and discussed for the remaining variable agent characteristics of infertility, puberty, and birth cost. 
 
\subsubsection{Sensitivities of Surplus Economies}

The sensitivities of population dynamics in surplus economies to small and large birth costs holding infertility at five and puberty at one are given in Figure 2. This figure shows clearly how the birth cost parameter affects the intrinsic growth rate. These large birth cost economies will play an interesting role in the later studies on inequality. The population dynamics at steady state are flat for high birth costs and move through the various regimes of constant level; and damped, steady, and chaotic magnitude periodic oscillations for the lower birth costs.

\begin{center}
	\centering
	\includegraphics[angle=-90,scale=0.8]{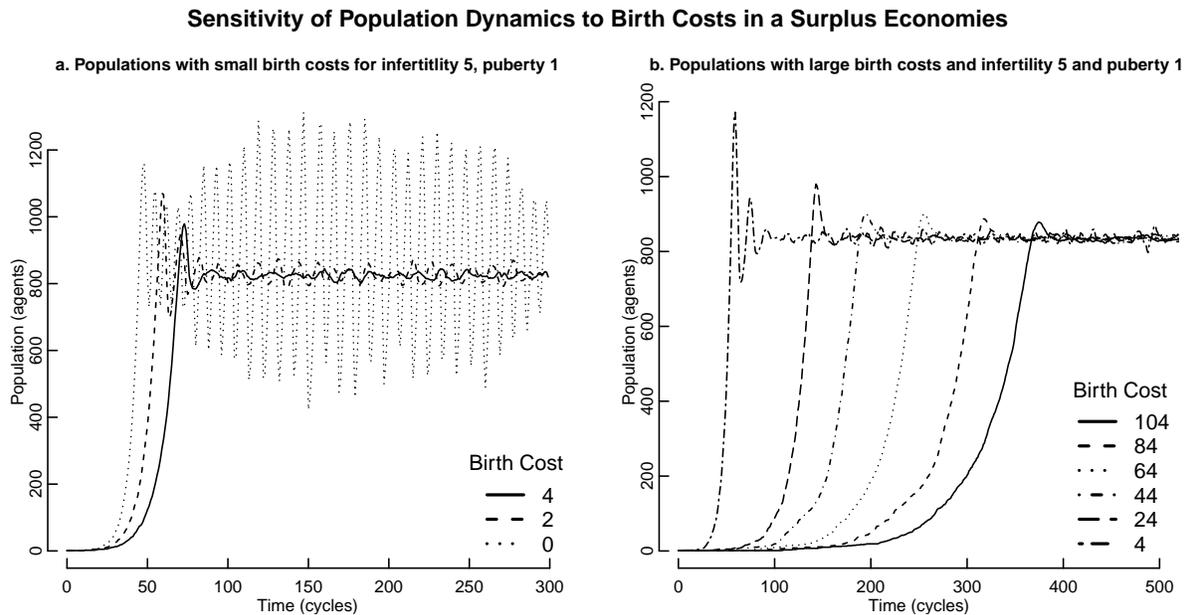}
	\captionof{figure} {a.) The population trajectories of surplus economies with various small birth costs (a) and large birth costs (b) for constant infertility ($f=5$) and puberty ($p=1$). The sensitivity of the trajectories to decreasing birth cost can be seen both in increasing rates of growth to carry capacity, and in the increasing magnitude of oscillations.} 
\end{center}

The sensitivity of the population dynamics to the infertility parameter ($f$) with puberty and birth costs both held at zero are given in Figure 3(a). These sensitivities to puberty with infertility held constant at 1 and birth cost 0 are given in Figure 3(b). Both figures show the same regimes of constant level; and damped, steady, and chaotic magnitude periodic oscillations as seen for the previous birth cost sensitivities. These  sensitivities also give rise to the first examples of a population going extinct. Individual-based models provide significant and unique insights into extinctions.

\begin{center}
	\centering
	\includegraphics[angle=-90,scale=0.8]{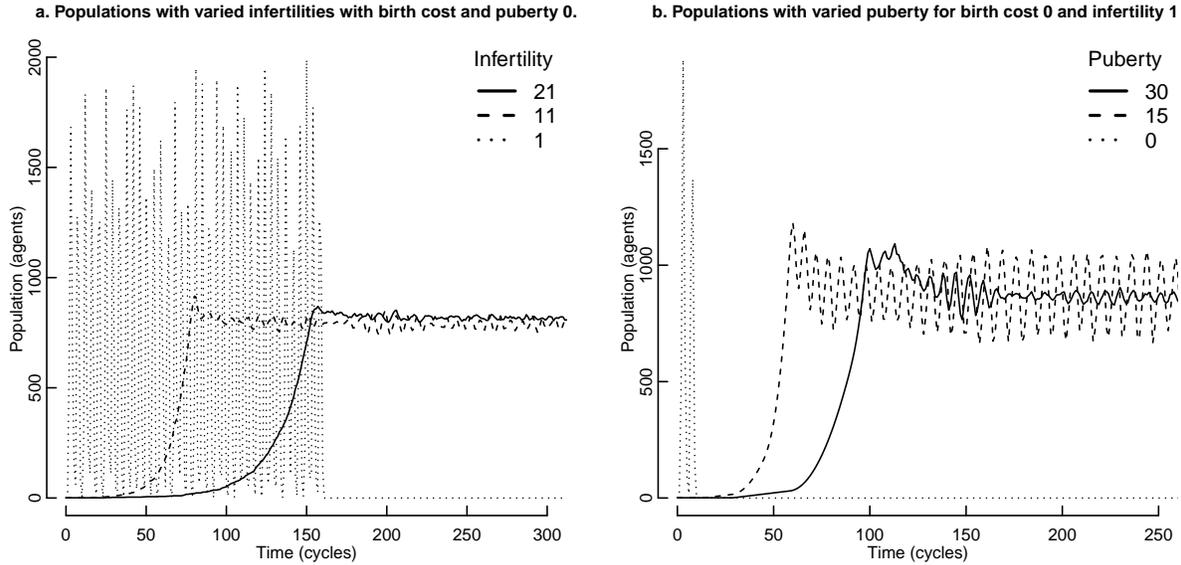}
	\captionof{figure} {a) The population trajectories of surplus economies with varied infertility for constant puberty ($p=0$) and birth cost ($bc=0$). b) The population trajectories of surplus economies with varied puberty for constant infertility ($f=1$) and birth cost ($bc=0$). Decreasing values of infertility or puberty increase the growth rate towards carry capaciity and increase the magnitude of oscillations. Both these graphs show that at high enough growth rates, chaotic extinctions occur [Section 3.2]. } 
\end{center}

While population trajectories attain their steady state modes quite quickly; the individual, local characteristics of the population, such as age or surplus resources, may not. In fact, these individual characteristics may take as long as another order of magnitude of time to reach steady state. These transition periods turn out to be very important for many individual level measurements of population characteristics and are treated in detail in Section 4.0.

\subsubsection{Sensitivities in Subsistence Economies}

Population growth in a subsistence economy is a more challenging activity for the individuals. By definition, the amount of resources consumed during a cycle is equal to the maximum available from a landscape cell. There are no resources remaining to support replication unless the birth cost is zero. Figure 4 below shows the sensitivity of population growth to infertility and puberty. All the now familiar regimes are represented here as well.

\begin{center}
	\centering
	\includegraphics[angle=-90,scale=0.8]{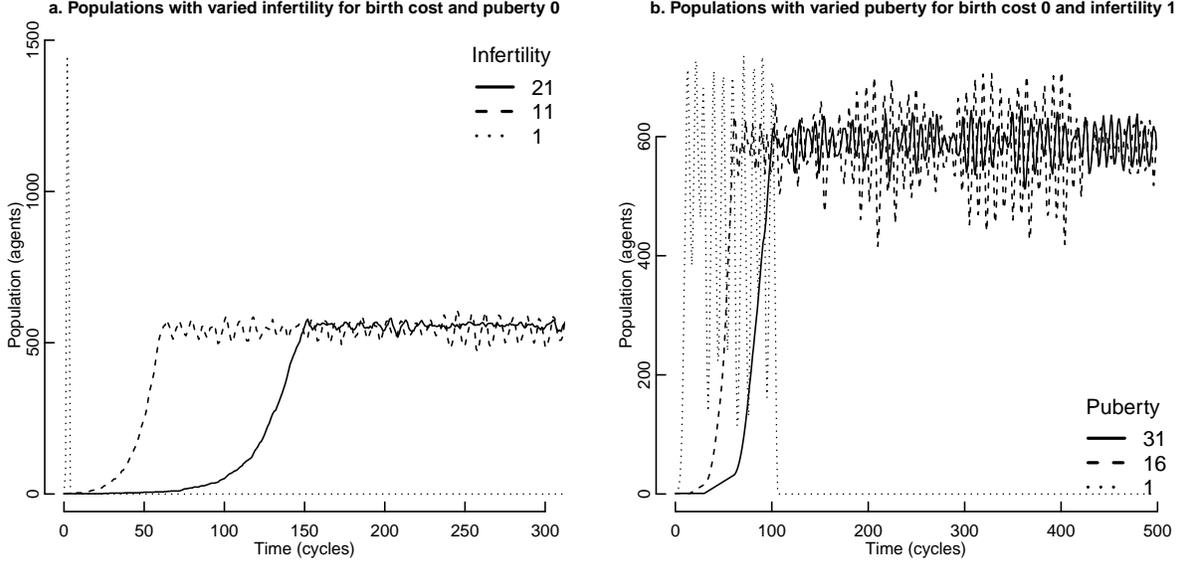}
	\captionof{figure} {a) The population trajectories of subsistence economies with varied infertility for constant puberty ($p=0$) and birth cost ($bc=0$). b) The population trajectories of subsistence economies with varied puberty for constant infertility ($f=1$) and birth cost ($bc=0$). These trajectories show similiar sensitivities as the surplus economies (Figure 8) but with a lower actual carry capacity and somewhat more chaotic magnitudes of the oscillations. Chaotic extinctions [Section 3.2] are present in both graphs and the f1p0bc0 extinction is the first example of a density-limiting high growth extinction [Section 3.1].} 
\end{center}

For the subsistence economies, extinction events are more common. Figure 4(a) shows an immediate extinction due to a very high intrinsic growth rate. Figures 4(b) and 3(a) provide examples of extinctions resulting from chaotic magnitude periodic oscillations.

\subsection{Global Modeling from Mathematical Biology}

The fields of mathematical biology and ecology developed equation-based continuum modeling of single species populations, models both continuous and discrete (Murray, 2002; Kot, 2001). A continuous homogeneous model of single species population $N(t)$ was proposed by Verhulst (1838):

\begin{equation}
	\frac{dN(t)}{dt}=rN(1-\frac{N}{K})
\end{equation}
where $K$ is the steady state carry capacity, $t$ is time, and $r$ is the intrinsic rate of growth. This model represents self-limiting, logistic growth of the population. This macroscopic model of a continuous, homogeneous population is quite descriptive and allows the exact solution
\begin{equation}
	N(t)=\frac{K}{[1+(\frac{K}{N_{0}}-1)e^{-rt}]}
\end{equation}
where $N_{0}$ is the initial population. 

Table 4 provides the implied intrinsic growth rates $r_{i}$ with standard deviation and implied carry capacity $K_{i}$ based on fitting the solution of the Verhulst Model (Equation 8) to simple surplus and subsistence economies with representative infertility, puberty and birth cost parameters. The stochastic nature of the initial growth of slow-growth economies generated a significant dispersion of the time of maximum growth but not the rate. The mean of the individual fits for each of one hundred population trajectories was selected as the better method to imply the intrinsic growth rate. Producing a fit for all the data at once provided a poor estimate of the maximum rate of growth. To confirm this choice, the initial population was increased from 2 agents up to 20 agents, reducing the time delay in achieving maximum growth rates. As the initial population approached 20 agents, the implied intrinsic growth rate of the "fit" approach converged towards the "mean" calculated value.
\begin{table}[h!]
	\centering
	\begin{tabular}{|c|c|c|c|c|c|c|}
		\hline
		Simple Economy & mean $r_{i}$ & mean $K_{i}$ & mean window & fit $r_{i}$ & fit $K_{i}$ & Plot \\ 
		\hline
		f85p1bc0m3 & $0.0256\pm3.62\%$ & 845 & 2,000 & $0.0175\pm0.40\%$ & 851  &  Fig 5(a)\\
		f10p1bc40m3 & $0.0323\pm0.577\%$ & 829 & 1,000 & $0.0322\pm0.12\%$ & 829 &  Fig 5(a)\\
		f10p1bc0m3 & $0.231\pm3.52\%$ & 878 & 200 & $0.169\pm0.686\%$ & 863 &  Fig 5(b)\\
		f5p1bc0m3 & $0.477\pm3.41\%$ & 881 & 200 & $0.393\pm3.31\%$ & 879 &   Fig 5(b)\\
		\hline 
		f20p1bc0m4 & $0.0968\pm4.90\%$ & 586 & 500 & $0.0723\pm0.641\%$ & 586 &  Fig 5(a)\\
		f10p1bc0m4 &  $0.181\pm5.39\%$ & 590 & 200 & $0.1467\pm0.691\%$ & 591 & Fig 6(a)\\
		f5p1bc0m4 & $0.364\pm5.68\%$ & 572 & 200 & $0.286\pm1.59\%$ & 570 &  Fig 6(a)\\
		\hline
	\end{tabular}
	\caption{Simple Economies fit to the Verhulst Model (Equation 7). One hundred runs of each simple economy were used for a least square fit to the Verhoulst parameters due to the stochastic nature of the initial growth from two agents. The mean window was taken to be proportional to the time to achieve the population's carry capacity to prevent overweighting the fit for the growth parameters by a long run at carry capacity. The "mean" of each individual implicit growth rate preserved the time correlation of the data whereas the "fit" values solved all the runs at once.}
	\label{Table 4:}
\end{table}

Figures 5 and 6(a) show how the Verhulst Model with these implied parameters compare to the actual population trajectories of these simple economies.

\begin{center}
	\centering
	\includegraphics[angle=-90,scale=0.8]{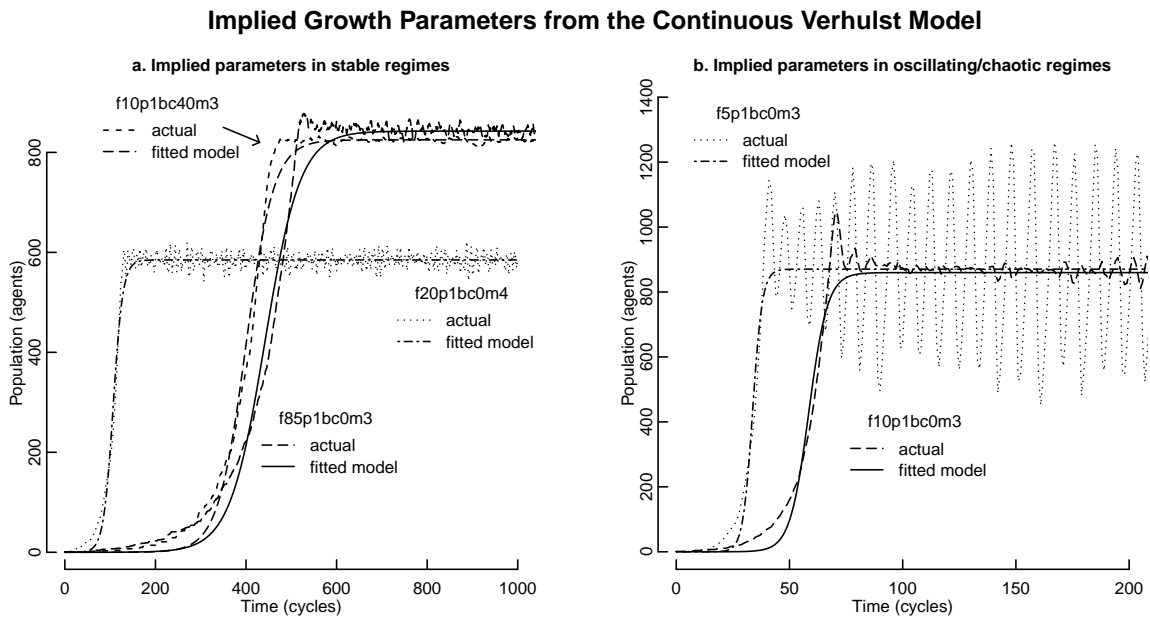}
	\captionof{figure} {Implied growth parameters for (a) representative stable, simple economies (f85p1bc0m3, f20p1bc0m4 and f10p1bc40m3) and for (b) representative oscillating/chaotic, simple surplus economies (f10p1bc0m3 and f5p1bc0m3). These implied parameters are least-squares fits to the continuous Verhulst Model (Equation 7) of population trajectories generated by the simple economic model. While these parameters fit the intrinsic growth rate of Verhulst Model and provide insights into the relationship of the model's growth parameters to the intrinsic growth rate, the Verhulst model does not admit oscillating behaviors as seen with the simple economic model.} 
\end{center}

As can be seen from these figures, while the Verhulst Model fits the initial phase of growth well, it does not model the oscillating population levels at the higher rates of intrinsic growth. The types of periodic and chaotic oscillations seen in these simple economies are often generated by population models that are discrete, that contain time delays, or both (Liz, 2014)

\begin{center}
	\centering
	\includegraphics[angle=-90,scale=0.8]{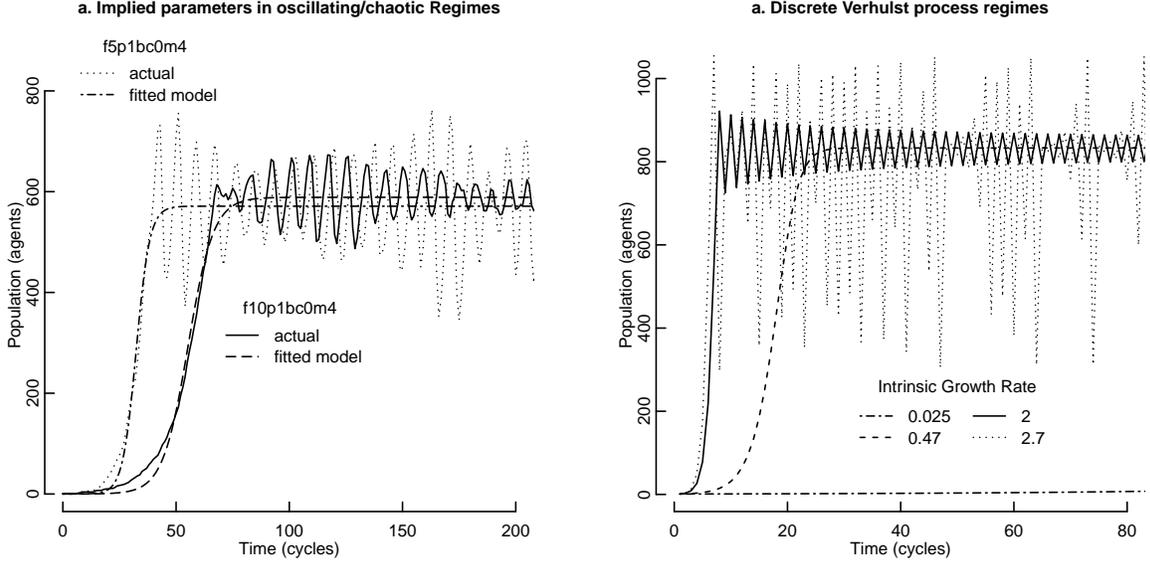}
	\captionof{figure} {a) Implied growth parameters for representative oscillating/chaotic, simple subsistence economies (f10p1bc0m4 and f5p1bc0m4). These implied parameters are least-squares fits to the continuous Verhulst Model (Equation 7) of population trajectories generated by the simple economic model. The Verhulst Model with the implied growth parameters from the simple econommic model matches the growth rate well but is not capable of modelling the oscialltions generated by the simple economic model. b) Representative critical regimes of the discrete Verhulst process (Equation 10). The discrete Verhulst process, while representing the discrete version of the continuous Verhulst Model, is capable of generating oscillations with decaying, steady, and chaotic magnitudes, similar to what emerges from the simple economic model.} 
\end{center}
Researchers in the fields of biology and ecology have used these discrete and delayed population models to handle, for example, species that have no overlap between generations (Murray, 2002) or have specific breeding seasons (Kot, 2001). Writing the Verhulst model (Equation 7) as a difference equation:
\begin{equation}
N(t+1)-N(t)=rN(t)[1-N(t)/K]
\end{equation}
and rearranging yields the discrete logistic growth equation, often referred to as a discrete Verhulst process (May, 1974):

\begin{equation}
	N(t+1)=[1+r-\frac{rN(t)}{K}]N(t)
\end{equation}
where $K$ is the carry capacity treated as free parameter. The discrete Verhulst population trajectories have different regimes that are quite similar to the trajectories of simple economies and very different from the continuous model. Figure 6(b) plots the various dynamic regimes generated by critical intrinsic growth rates for the discrete Verhulst process: constant level; and damped, steady, and chaotic magnitude periodic oscillations. The intrinsic growth rates representative of each of the regimes in the Verhulst process (Equation 10) are much greater than the implied growth rates (Table 4) from the continuous Verhulst model (Equation 7). Since the aggregate behavior of the simple economic model and the implied growth rates were in agreement, the discrete Verhulst process may still be missing a significant factor.

Once a landscape cell's resources have been consumed, for a growth rate of one resource per cycle, it takes four cycles for a complete restoration of the cell's resources. A subsistence economy requires four cycles of growth to match the agent's metabolism while surplus economies require only three cycles. To account for similar delays in animal populations, Hutchinson (1948) modified the Verhulst process (Equation 10) as: 
\begin{equation}
	\frac{dN(t)}{dt}=rN(t)(1-\frac{N(t-\tau)}{K})
\end{equation}
by incorporating an explicit time delay $\tau$ in the self-limiting term. The resulting discrete-delayed logistic equation (Wright, 1955), often referred to as the Hutchinson-Wright equation, (Kot, 2001) is then

\begin{equation}
N(t+1)=[1+r-\frac{N(t-\tau)}{K}]N(t)
\end{equation}

Figure 7(a) plots the population dynamics generated by this Hutchinson-Wright equation with $\tau$ equal to three and the implied growth rates and carry capacities from the continuous Verhulst equation for the simple surplus economies (Table 4). Figure 7(b) plots the same results for $\tau$ equal to four and the implied parameters for the simple subsistence economies (Table 4). Magnitudes of intrinsic growth rates for each of the dynamic regimes generated by the Hutchinson-Wright process show better agreement with implied growth rates for the simple economies by matching the delay with replenishment times.

\begin{center}
	\centering
	\includegraphics[angle=-90,scale=0.8]{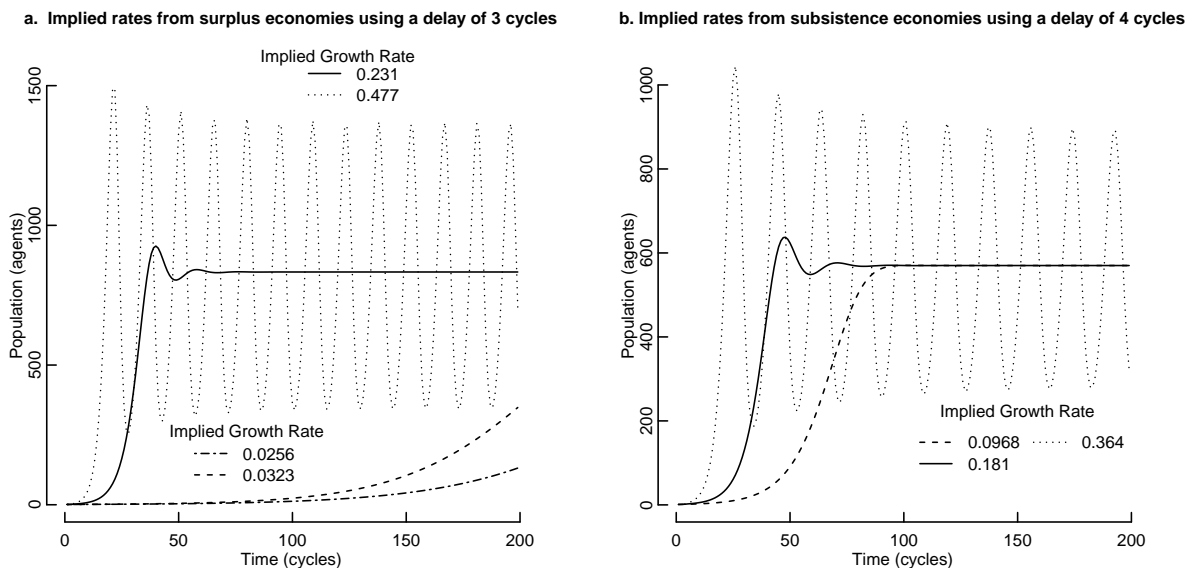}
	\captionof{figure} {Implied intrinsic growth parameters for representative simple economies (Table 4) applied to the descrite Hutchinson-Wright process (Equation 12). a) Surplus economies applied with a delay parameter ($\tau$) of 3 cycles and b) subsistence economies applied with a delay parameter of 4 cycles. The critical intrinsic growth rates and delay parameters of the Hutchinson-Wright process show good agreement with the intrinsic growth rates implied from fitting the population trajectores generated by the simple economic model to the continuous Verhulst model (Equation 7) when using the appropiate delay parameter for surplus and subsistence landscape growth rates.} 
\end{center}

\subsection{Discussion}
The population growth of simple economic models have generated distinct dynamic regimes: initial exponential population growth; constant level; damped, steady, and chaotic magnitude periodic oscillating population trajectories; and extinction events. A comparison of these dynamics with the family of single species logistic equations validates these computational results to well established global population models of biology and ecology and provides insight into how parameters of the simple economic model map to intrinsic growth rates of the equation-based models of mathematical biology.

\section{Extinction}
Studies of extinction events are aided by the spatial-temporal resolution provided by individual-based models. In particular, for these simple economies, extinction events illustrate both the importance of a precise understanding of the simulation loop as well as the benefits of spatial resolution of individuals. This section will look at both immediate extinctions due to very high intrinsic growth rates and delayed extinctions resulting from regimes with chaotic trajectories. As part of this discussion, measures of self-limiting growth and local versus global resource availability are developed.
   
\subsection{Local Damping of High Growth Extinctions}

As the previous sensitivity studies have shown, certain combinations of model parameters result in extinctions. With parameters set to maximum intrinsic growth in a subsistence economy (f1p0bc0m4), the population ramps up quickly and within five cycles goes extinct for all one hundred separate runs. Figure 8(a) plots the population trajectories of these hundred separate runs. 

\begin{center}
	\centering
	\includegraphics[angle=-90,scale=0.8]{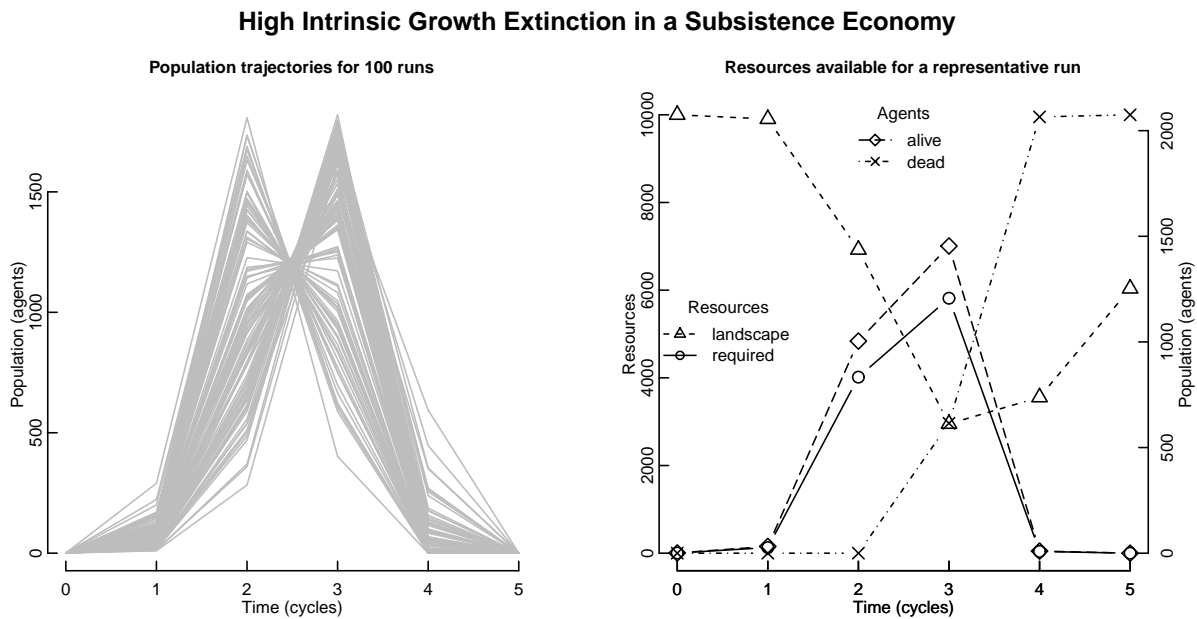}
	\captionof{figure} {a) The population trajectories for 100 runs of a high intrinsic growth simple subsistence economy (f1p0bcom4). All the runs are extinct by the fifth cycle and the initial growth is density limited for the first two cycles. b) The resources available versus resources required by the current population for a sample run taken from the 100 runs. The available landscape resources are insufficent to support the population in cycle 3 and extinction is predicted by these global measures.} 
\end{center}
Figure 8(b) shows the depletion of the resources within the landscape below that required by the current population for one of these runs at cycle 3. The last few agents alive in cycle 4, even after surviving the resulting large die off, expire due to an apparent lack of local resources though the global resources would now support the remaining population. This phenomenon of the local landscape being insufficient to support a population even though the total landscape resources are sufficient is common in most observed extinctions. A measure of these locally available resources is developed in Section 3.2 on chaotic extinctions. First, the density limitation of initial growth of this simple economy is examined.

Starting with one agent and an initial landscape maxed out at four resources in every cell, there are many births during the very first cycle. With puberty set to 0 cycles, each new agent has the opportunity to immediately reproduce in the cycle in which it was born. What prevents the landscape from completely filling immediately is the requirement that, for an agent to reproduce, one of the four neighboring cells of that agent be unoccupied. Following the agents' movement for resources and then reproduction, it can be seen in Figure 9(a) how one trajectory hits the no-empty-neighbor condition after 30 births during the first cycle. Figure 8(a) shows that all the one hundred trajectories hit this local density-limited growth within 200 births. With such fecundity, even with this density-limited growth, it is not surprising that the population exceeds the theoretical carry capacity within three cycles and goes extinct within five cycles for all the runs. A measure of this local density limitation, a crowding factor $C_{j}^f$, is defined for the reproducing agent in $c_{j}$ as
\begin{equation}
C_{j}^{f}=\sum_{i=1}^{V_{j}(1)}c_{i}^{empty}
\end{equation}
where $V_{j}(1)$ are the von Neumann neighbors at a distance of 1 from cell $c_{j}$, and $c_{i}^{empty}$ is a Boolean function with value zero if $c_{i}$ is occupied by another agent, otherwise it equals one, yielding a count of the number of empty, neighboring cells. Figure 9(b) plots the crowding factor for each reproducing agent from Figure 9(a), demonstrating that the lack of a empty von Neumann neighbor ended the first cycle.

\begin{center}
	\centering
	\includegraphics[angle=-90,scale=0.8]{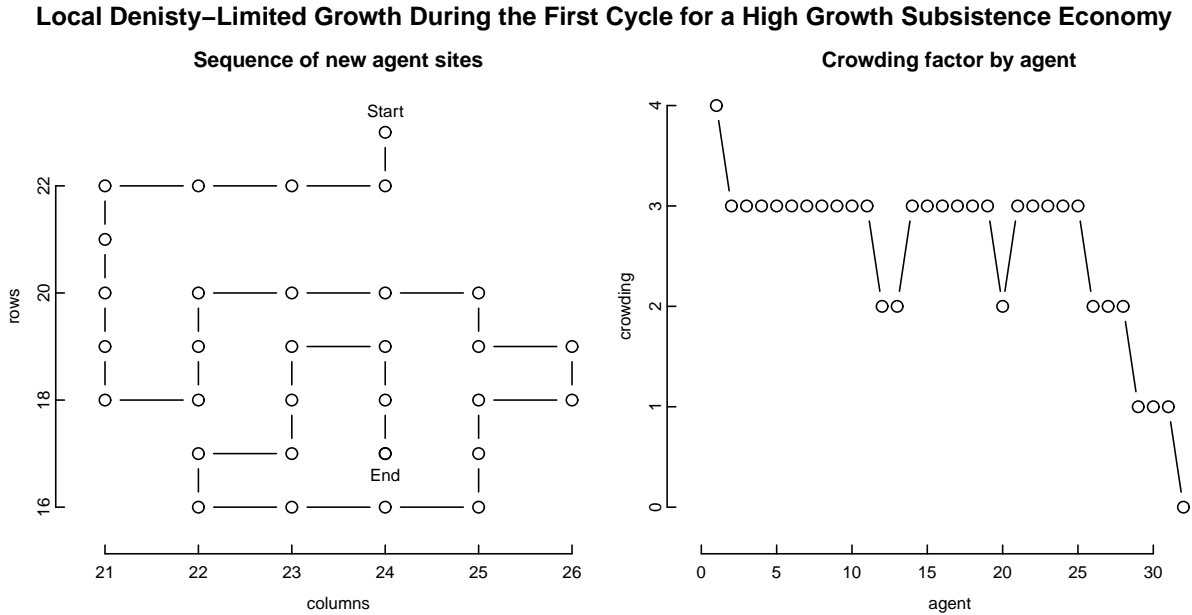}
	\captionof{figure} {a) New agent birth sites during the first cycle of high growth for a simple subsistence economy (f1p0bc0m4). The puberty value of zero allows a new agent to immediately enter the action queue. The sequence ends when there are no empty sites ($V_{j}(1)$) available for the latest agent in cell $j$. b) The crowding factor (Equation 13) for the sequence of new agents shown in (a). When the crowding factor reaches zero, the growth is density-lmited and the cycle ends.} 
\end{center}

\subsection{Chaotic Extinctions}

Setting the puberty to one cycle significantly reduces the intrinsic growth rate and delays but does not necessary avoid extinction. Figure 10(a) presents a histogram of the extinction time of all thousand runs of such a simple economy (f1p1bc0m4), with only a few outlying runs extending past 800 cycles. Figure 10(b) plots the chaotic population trajectory for the run that lasted the longest in this set of 1000.

\begin{center}
	\centering
	\includegraphics[angle=-90,scale=0.8]{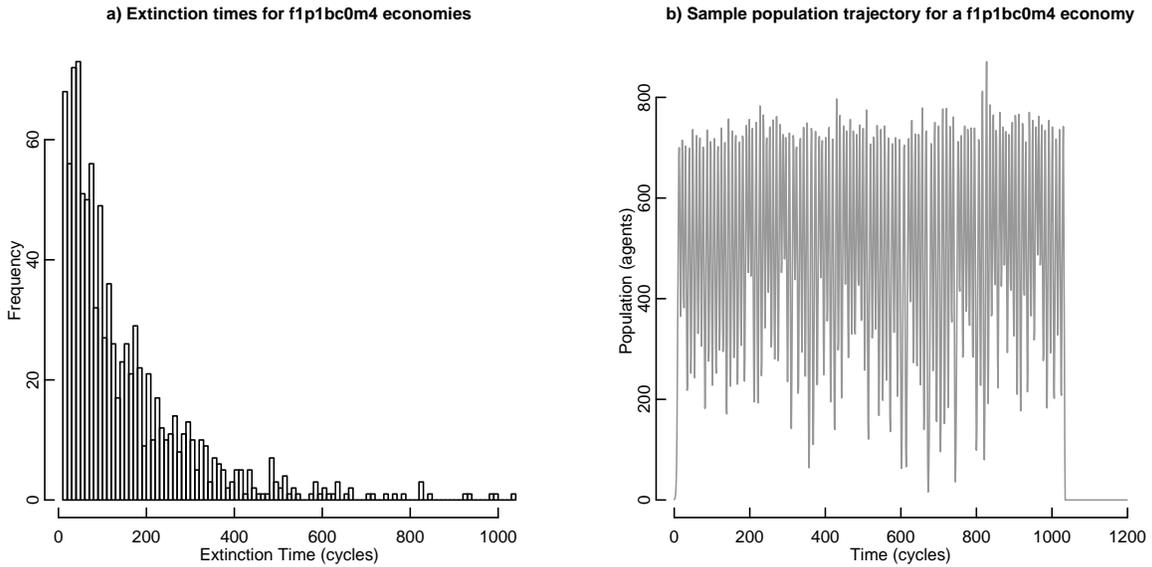}
	\captionof{figure} {a) Histogram of extinction times for a 1,000 runs of a chaotic subsistenace economy (f1p1bc0m4). Most extinctions occur within the first few hundred cycles, with the longest lasting run reaching over 1,000 cycles. b) The population trajectory for the longest lasting run showing the oscillations of chaotic magnitude. All the trajectories in (a) display this chaotic behavior resulting in stochastic extinction times for the same simple economy.}
\end{center}
Of these 1000 f1p1bc0m4 trajectories, a trajectory with an extinction time of 162 cycles is selected for more detailed examination of this different mechanism of extinction than seen in Section 3.1. The global or macroscopic measures of population and landscape resources of this selected trajectory are given in the Figure 11(a).
\begin{center}
	\centering
	\includegraphics[angle=-90,scale=0.8]{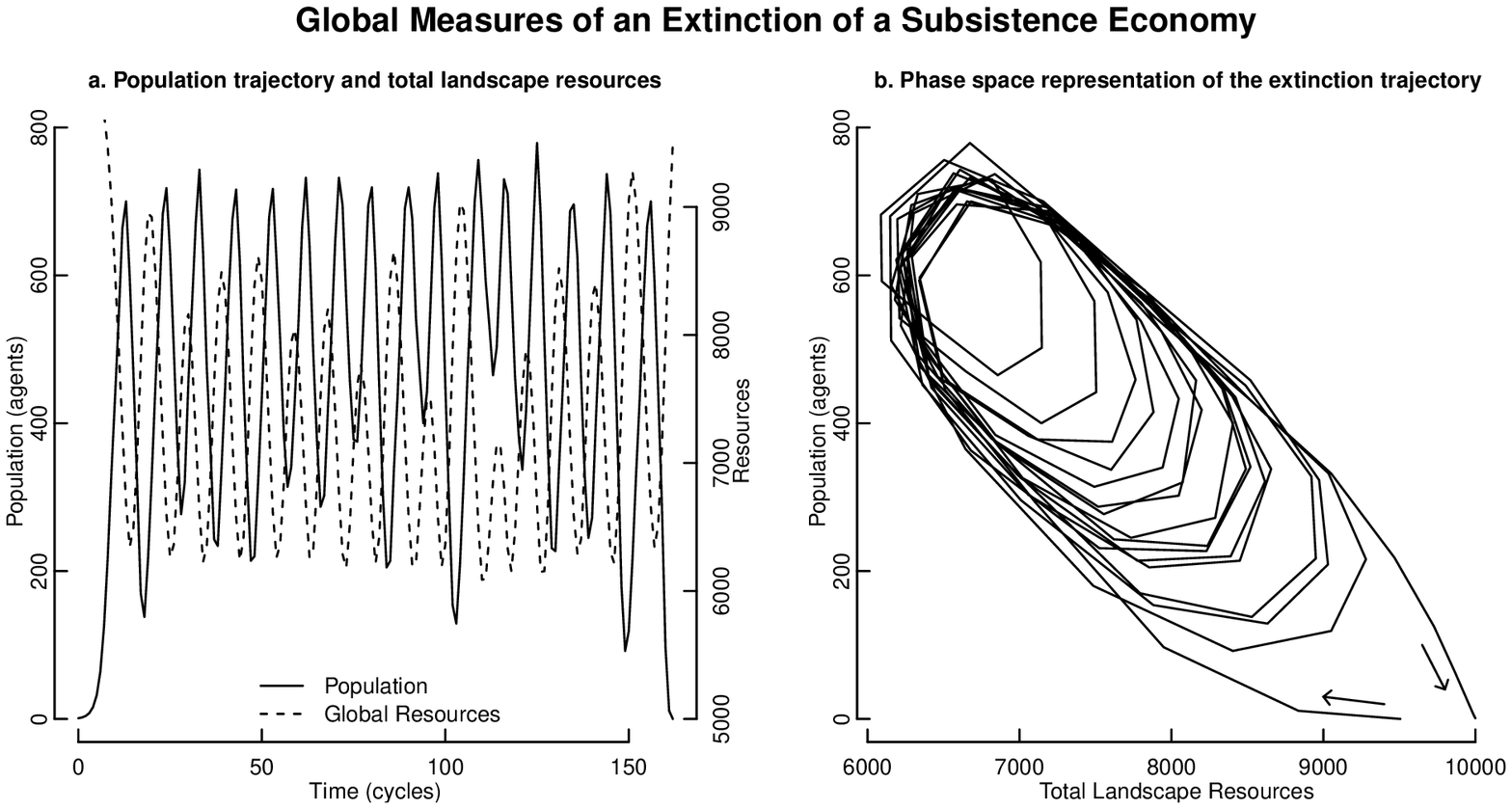}
	\captionof{figure} {a) The coupled, chaotic magnitude oscillations of levels of population and total landscape resources for a simple, subsistence econmy (f1p1bc0m4). While the total landscape resources always remain well above the level necessary to keep a good fraction of the population alive, the extinction nonetheless occurs. b) Representation in phase space of the levels of population and total landscape resources about the f1p1bc0m4 attractor. This visualization shows clearly that the total landscape resources never dip below 6,000 resources and the extinction occurs at a maximum in toal landscape resources. }
\end{center}
Though the population fluctuates wildly, and appears to be oscillating chaotically, it manages to avoid extinction for 15 separate minimums over 160 cycles. Furthermore, as Figure 11(a) shows, the global resources available in the landscape are always sufficient to support at least a small population. Figure 11(b) plots this chaotic trajectory in the phase space of the population and global resources about the attractor. This extinction must, therefore, be driven by a local, spatial dynamic rather than global measures. 

Defining $R_{j}^{l}(t)$ to be the local measure of resources within range of an agent occupying $c_{j}$ as
\begin{equation}
R_{j}^{l}(t)=\sum_{i=1}^{V_{j}(v)}r(c_{i},t)
\end{equation}
where $V_{j}(v)$ are all the von Neumann neighbors within movement/vision range $v$ of cell $c_{j}$, and $r(c_{i},t)$ are the currently available resources in $c_{i}$ at time $t$. Averaging over all the agents $A(t)$ defines an averaged local measure of resources available to the population $\bar R_{total}^{l}$  as
\begin{equation}
\bar R_{total}^{l}(t)=\sum_{i=1}^{A(t)}R_{j}^{l}(t)/A(t)
\end{equation}
Figure 12(a) provides $\bar R_{total}^{l}$ and the global population versus time. $\bar R_{total}^{l}$ is a predictor of a pending extinction in contrast to the total resources available [shown in Figure 11(a)] which does not recognize an impending extinction event.
\begin{center}
	\centering
	\includegraphics[angle=-90,scale=0.8]{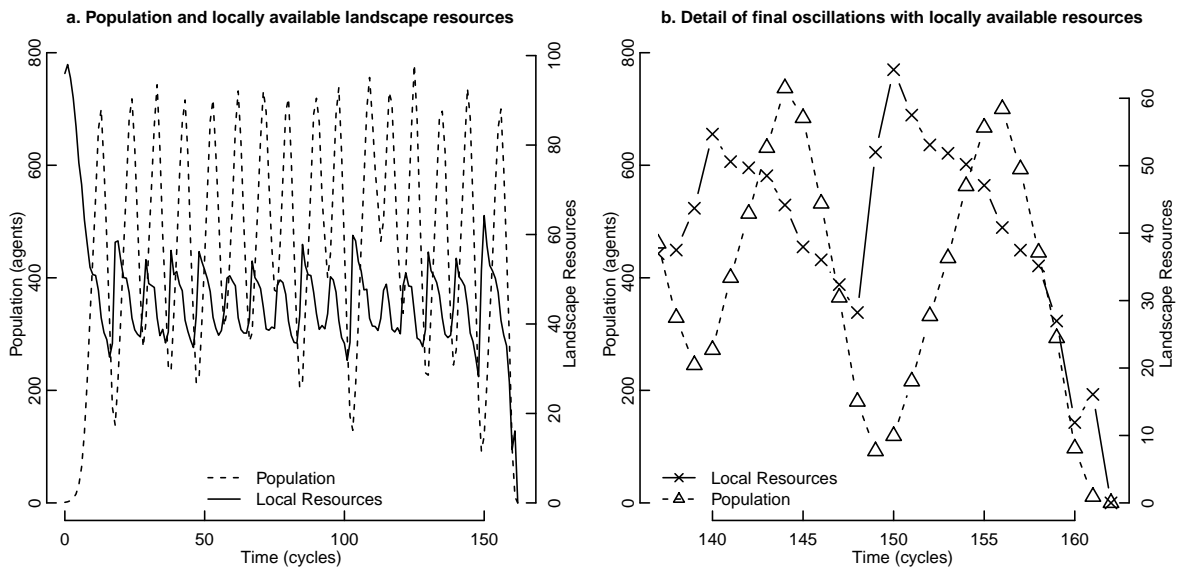}
	\captionof{figure} {a) The population trajectory and averaged locally available landscape resources ($\bar R_{total}^{l}$) for a simple subsistence economy (f1p1bc0m4). b) Detail of the final two oscillations leading to extinction. For the last three cycles, the averaged locally available resources were never enough to support the dwindling population though the global resources, shown in Figure 11, had fully recoverd. The last remaining agents perish in a local desert with plentiful resources too far away.}
\end{center}
Figure 12(b) provides greater detail of the final oscillations that results in the extinction event. Spatial-temporal characteristics of this extinction event are examined in further detail in the next section. 

\subsection{Microstructure of Extinctions}

To help understand the details of how an extinction event is driven by non-equilibrium, local, spatial, and stochastic effects, Figure 13 provides a spatial visualization for four cycles of the agent locations and the resources available for the last two minimums of the run detailed in Figures 11 and 12. Figure 13(a) and (b) show the penultimate oscillation and recovery. 
\begin{center}
	\centering
	\includegraphics[angle=-90,scale=0.8]{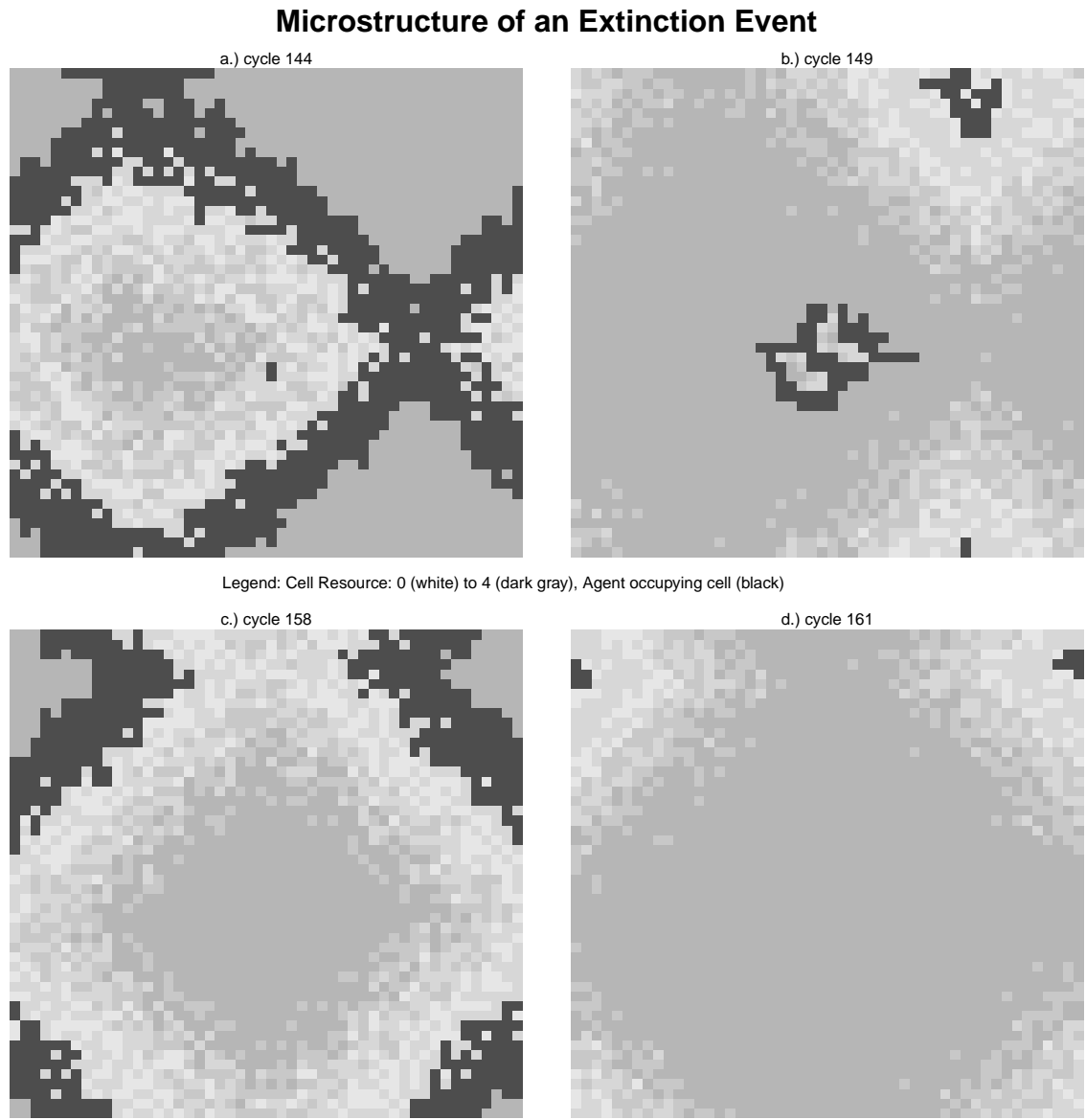}
	\captionof{figure} {a) When the penultimate wave of agents pass through the rich landscape, a couple agents manage to survive behind the wave. b) As the last agents of the wave perish in a local desert, the next generation blooms. c) For this final wave, no agents survive behind the wave as it passes. d) The last cycle with the remaining agents all in a local desert and about to perish. This extinction event is driven by non-equilibrium movement dynamics, and spatially local resource densities. Together these factors produce a highly stochastic process leading to certain extinction but at an unpredictable time.}
\end{center}
At the agent level, the agents form a cylindrical wave of resource depletion moving from a starting small group of agents at the center and ranging radially outward until all the resources have been consumed. Note that two agents in Figure 13(a) have survived in the center after the consuming wave moves out radially. Figure 13(b) shows the last remnants of that wave expiring at the edges as the next generation blooms from the agents that survived in the center. Thus the population as a whole survived this oscillation. Figure 13(c) shows the next oscillation, with the wave of agents flowing out from the center again but this time, for the first time, no agents in the center are able to survive the passing of the wave. Thus, Figure 13(d) shows the extinction of the population as the last remnants of the wave perish in a local desert with no agents left behind in the middle to seed the next generation. This extinction is purely a local, and spatial phenomenon since the total resources are as much or more than the numerous previous oscillations which avoided extinction. The extinction is also stochastic since the survival of a few agents in the center is likely but not certain.

\section{Resource Inequality}

In addition to the spatial resolution that is so important in understanding extinction events, additional measures of distributions of individual characteristics are available with this simple economic model. Three of these measures are the individuals' ages, surplus resources (wealth), and the deaths per cycle. This section will first investigate the sensitivities of the populations' mean age and inequality distributions to the model's parameters at steady state. Four representative surplus economies are then identified as models for further study into the individuals' distributions. Large differences in inequality due to the different growth parameters are highlighted and the causes of inequality in homogeneous (equal ability) agent populations in an homogeneous (equal opportunity) resource environment are considered. Emphasis is placed on the differences in these characteristics between steady state (equilibrium) and transient (non-equilibrium) conditions.

\subsection{Inequality Sensitivities to Agent Characteristics}

A proxy for measurement of inequality in populations often used and misused is the Gini Coefficient, $G$, defined as:
\begin{equation}
G = \mathbb{E}(|S^{'}-S^{"}|)/(2\mu)
\end{equation}
where $\mathbb{E}$ is the expectation operator, $S^{'}$ and $S^{"}$ are the surplus resources of different agents and $\mu$ is the mean of the surplus resources $\mu = \mathbb{E}(S)$ (Yitzhaki and Schechtmann, 2012).

Computationally, Equation 16 was implemented as:
\begin{equation}
G=\frac{\sum_{i=1}^{A}\sum_{j=1}^{A}|S_{i}-S_{j}|}{2A\mu}
\end{equation}
where $i$ and $j$ are indices to run through the entire population $A$ of agents who have $S_{index}$ surplus resources. Though there are many problems with describing whole distributions with a single number, by computing the Gini Coefficient over the entire population rather than sampling the population, by not assuming a particular distribution, and by only using Gini Coefficients as relative measures for comparing populations of the same size, these inaccuracies can be minimized (Fontanari et. al., 2018). The Gini Coefficient is particularly suspect when used for comparisons of populations that are undergoing large oscillations in size over time (Talib, 2015).

In order to select appropriate model economies that will generate various inequality distributions, sensitivities of the mean age and inequality (as imperfectly expressed by the Gini Coefficient) to the growth parameters are computed. Care must be exercised to ensure the specific population has reached a steady state for all its measurement parameters, not just population size. While the population will hit its actual carry capacity quite quickly, the mean age of the population can take from one hundred to one hundred thousand cycles to reach its steady state.

Figure 14(a) plots the sensitivity of these distributional measurements to birth cost, and Figure 14(b), the sensitivity to puberty. The large variance in Gini Coefficient for repeated runs in the same configuration underscores the imperfect value of this metric at a specific cycle in oscillating and/or chaotic regimes.

\begin{center}
	\centering
	\includegraphics[angle=-90,scale=0.8]{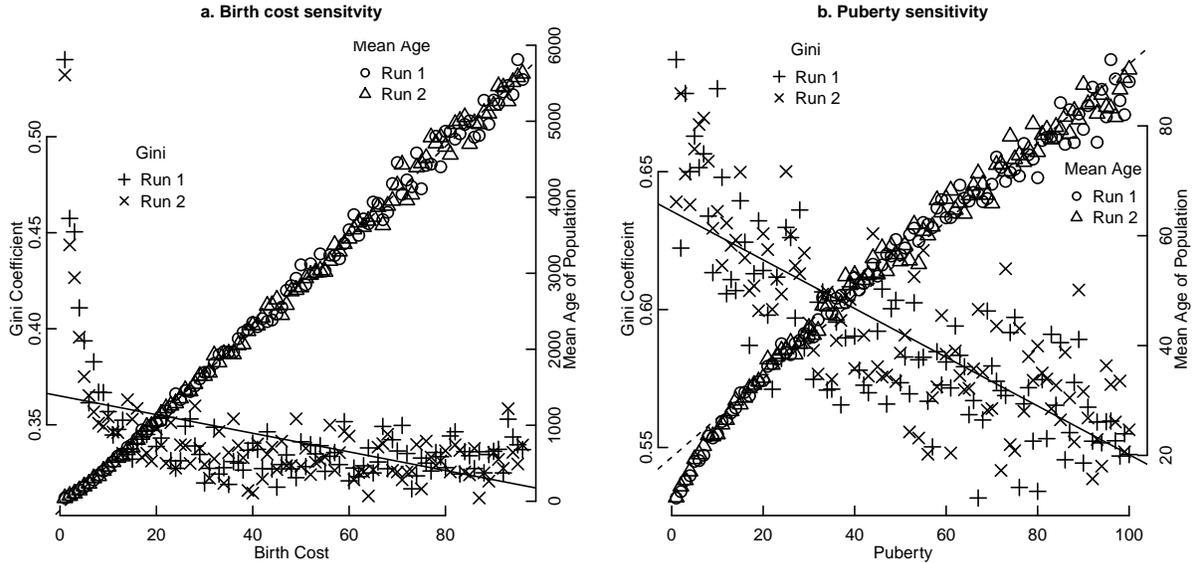}
	\captionof{figure} {a) The sensititivty of the mean age and Gini Coefficient to birth cost with constant infertility ($f=10$) and puberty ($p=1$) at the end of the 130K to 150K cycles steady state window.  The addition of birth costs to this simple f10p1 economy immediately and significantly reduces the inequality and holds it at these low levels as the birth costs continue to increase. The effeccts of increasing birth costs on the mean age of the population, in contrast, are quite linear. b) The sensititivty of the mean age and Gini Coefficeint to puberty with birth cost ($bc=0$) and infertility ($f=10$) at the end of the 10k to 20K cycles steady state window. Increasing puberty reduces inequality and increases mean age for the f10bc0 simple economy though the Gini Coeficient measurements have a large variance, suggesting a chaotic regime. }
\end{center}

Figure 15(a) plots the sensitivities of inequality and mean age to infertility. The variance of the Gini Coefficient at small infertility will be addressed below. Figure 15(b) presents the relationships of mean ages and Gini coefficients to the mean of the deaths per cycle ($\bar d_{c}$) over the steady state window for variations of the agents' growth characteristics. In Figure 15(b) it becomes clear that the slope of the relationship of mean age to mean deaths per cycle is significantly different for birth cost sensitivity than it is for puberty and infertility, This difference in slopes is the first indication that the individual dynamics of surplus economies with significant birth cost are substantially different than without birth cost. The sensitivity of the Gini Coefficient to these parameters is more complex. The change in slope of mean age for increasing deaths per cycle and the double slope reversal for the inequality measures indicate a phase change in the dynamics of these distributions. In general, across all these sensitivities, the higher the mean deaths per cycle, the lower the mean age and the higher the inequality. These tendencies are based solely on economies that have reached steady state and do not apply to non-equilibrium economies still in transition.

\begin{center}
	\centering
	\includegraphics[angle=-90,scale=0.8]{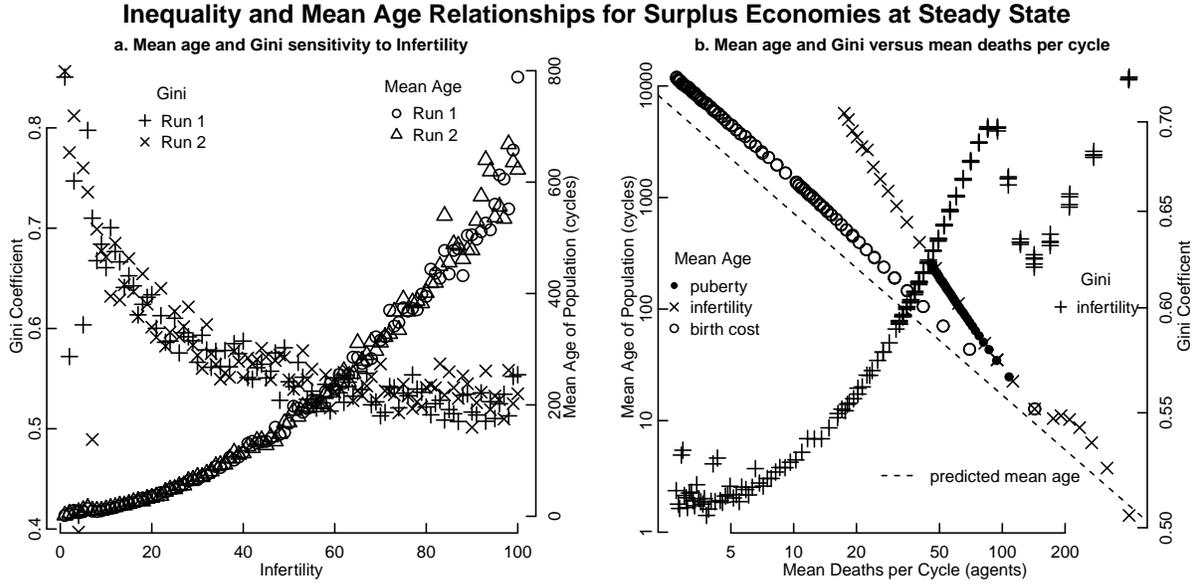}
	\captionof{figure} {a)  The sensititivty of mean age and Gini Coefficent measurements to infertility with constant puberty ($p=1$) and birth cost ($bc=0$) at the end of the 40K to 50K cycles steady state window. b) The relationship of mean age and Gini Coefficent to mean deaths per cycle over a broad range of model parameters (using the same steady state windows as before). A prediction of mean age based on probability of death per cycle (Equation 18) is plotted. Note the difference between the slope for mean age of birth cost sensitivity versus mean age sensititivies for puberty and infertitily on this log-log represntation.}
\end{center}

The predicted mean age model shown in Figure 15(b) is based on a Monte Carlo simulation of the steady state population $A$ run over time using the probability of an (mean) agent's death in any one cycle $P_{\bar d_c}$ to be given as
\begin{equation}
P_{\bar d_c} = \bar d_c /A
\end{equation}
This prediction shows good agreement with the mean age data for the birth cost sensitivity but gets the slope wrong for the puberty and infertility sensitivities and gives no indication of the change in linearity of the slope at high $\bar d_{c}$. The mean deaths per cycle metric, however, gives a good intuitive feel for what is happening to the population (level of carnage) with a particular configuration of growth parameters.

Figure 16(a) plots the relationship of the population's total wealth to mean age. The sensitivities of all three model growth parameters show increasing total wealth with increasing mean age, with similar and nearly constant slopes. The total wealth increases with increasing mean age even for increasing birth costs, which represent significant sunk costs. Figure 16(b) presents the relationship of the Gini Coefficient to the mean age of the population. A couple of complexities are apparent in Figure 16(b). First, two very different populations emerge with similar mean ages but significantly different inequality measures. Second, as mean age decreases to and below 10, an apparent phase transition occurs for both the total wealth and the inequality with two slope reversals.

\begin{center}
	\centering
	\includegraphics[angle=-90,scale=0.8]{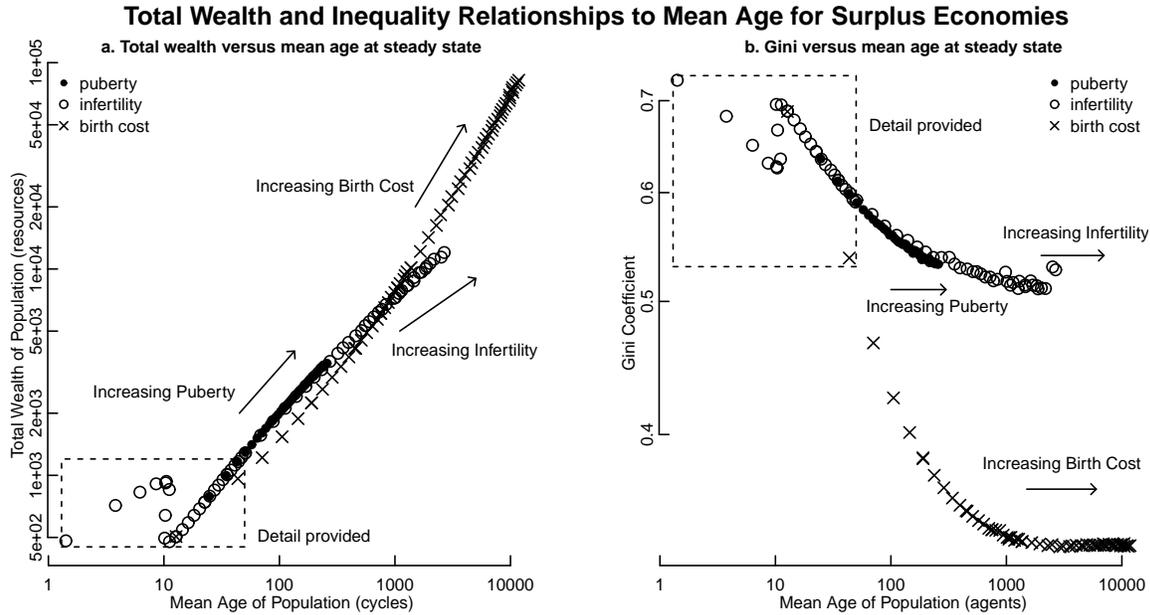}
        \captionof{figure} {a) The relationship of population's total wealth to mean age across a broad range of model parameters at steady state. The Details on the boxed area are provided by Figure 17(a). b.) The relationship of the Gini Coeficient to mean age across a broad range of model parameters at steady state. Details on the boxed area are provided by Figure 17(b). Increasing birth costs increase total wealth of the population in a similar fashion to other model parameters but have a strikingly stronger effect on inequality.}
\end{center}

Figure 17(a) provides detail of the measurement of total wealth and mean age and their one standard deviations from the highlighted (boxed) area in Figure 16(a). Figure 17(b) provides the detail of the measurements of the Gini Coefficient and mean ages and their one standard deviations from the highlighted (boxed) area in Figure 16(b). There is a significant increase in variance and a loss of one-to-one mapping of Gini Coefficient to mean age as the infertility passes through 10. This regime change is consistent with the Hutchinson-Wright process (Equation 12) with a delay of 3 cycles and an implied rate of growth increasing towards 0.477 as discussed in Section 2.5 and plotted in Figure 7(a). The means and standard deviations for Figure 17 were calculated over a steady state window of time from 10,000 to 20,000 cycles for mean ages less than 21 and from 200,000 to 225,000 cycles for mean ages greater than 20.

\begin{center}
	\centering
	\includegraphics[angle=-90,scale=0.8]{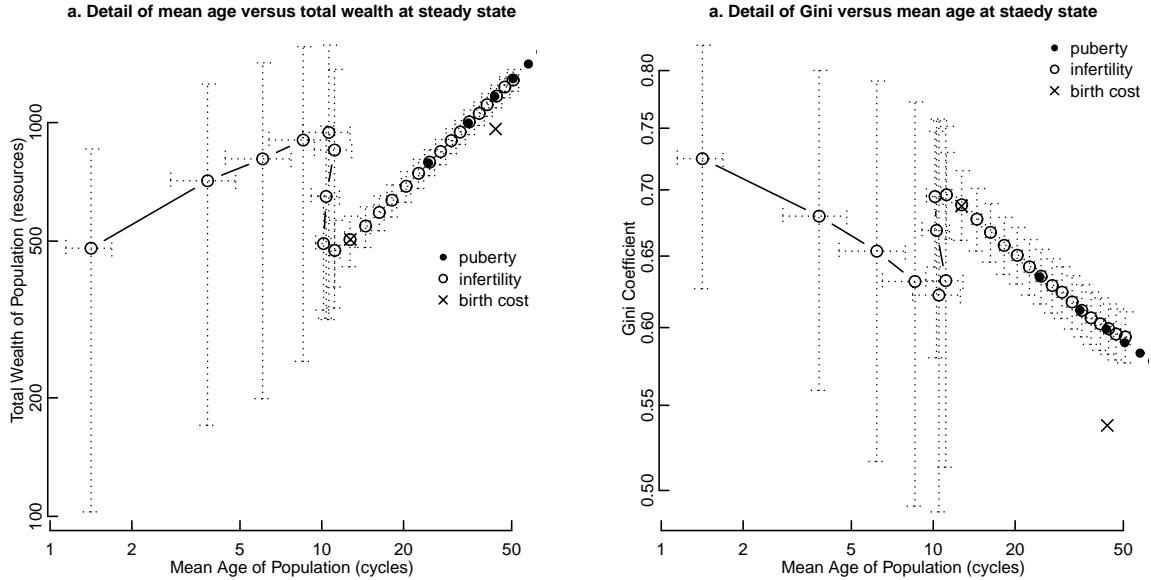}
	\captionof{figure} {a) The means and standard deviations of the total wealth and mean ages around the transition to high variance. The loss of one-to-one mapping of total wealth to mean age and the increase in variance indicate a transition to chaotic magnitude oscillations. b) The means and standard deviations of the Gini Coeficients and mean ages around the transition to high variance. The Gini Coefficeint sensitivity to mean age mirrors the transition behavior of the total wealth.}
\end{center}

\subsection{Model Surplus Economies}

From the sensitivities generated in the previous sections, three specific surplus economies with four population distributions are identified as model economies for more detailed study. Table 5 defines and labels these distributions taken from the model surplus economies. 

\begin{table}[h!]
\centering
 \begin{tabular}{|c|c|c|c|c|c|c|c|}
   \hline
Name & Infertility & Birth Cost & Growth Rate & Gini & Mean Age & Total Wealth & Steady State \\
(label) & ($1/P_{b}$) & (resources)&($cycles^{-1}$) & (-) & (cycles) & (resources) & (cycles)\\
   \hline
f85p1bc0m3\_50K & 85 & 0 & 0.0256 & 0.530 & 443 & 4861 & 35,000 \\
f85p1bc0m3\_6K  & 85 & 0 & 0.0256 & 0.777 & 1327 & 17,132 & 6,000* \\
f10p1bc0m3\_50K & 10 & 0 & 0.231 & 0.651 & 12.6 & 509 & 700 \\
f10p1bc40m3\_50K & 10 & 40 & 0.0323 & 0.332 & 2261 & 16,608 & 25,000 \\
   \hline
 \end{tabular}
 \caption{Definitions of Model Surplus Economies. (* not at steady state)}
 \label{Table 5:}
\end{table}

The trajectories of mean age of whole and elite populations for two model economies (f85p1bc0m3 and f10p1bc40m3) are given in Figure 18(a). The elite population consists of those agents whose wealth ranks in the top 10\% of the population. The f85p1bc0m3 economy has a very long relaxation time of 35,000 cycles to reach steady state. Of particular interest is the large peak of mean age of over 8,000 cycles for the elite population at a time of 10,000 cycles, dropping at a relative constant rate to a mean age of under 1,800 cycles at steady state. The whole population has a maximum mean age of 1,327 cycles at about 6,000 cycles. This population distribution at 6,000 cycles is the fourth model population (f85p1bc0m3\_6K) to be examined in detail though it is not at steady state. The second simple economy shown in Figure 18(a) is f10bc40m3\_50K with a relaxation time of 25,000 cycles.  This model economy displays a moderate relaxation time of 10,000 cycles to steady state with the highest steady state mean age of over 2,000 cycles. Figure 18(b) presents these measurements for the remaining model economy (f10p1bc0m3\_50K). This model economy has the shortest relaxation time of 700 cycles and mean ages under 20 cycles. 
\begin{center}
	\centering
	\includegraphics[angle=-90,scale=0.8]{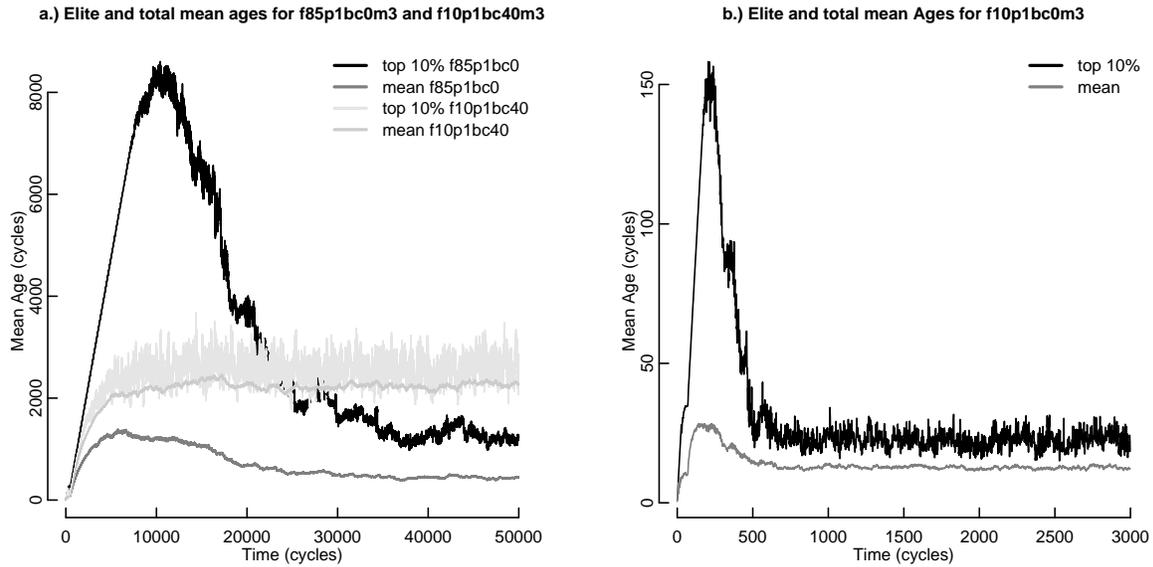}
	\captionof{figure} {a) Elite (top 10\%) and overall mean ages for surplus economies f85p1bc0 and f10p1bc40. b) Elite (top 10\%) and overall mean ages for surplus economy f10p1bc0. The large inequalities are generated during the population's growth to carry capacity and persist for periods much longer than the time to reach carry capacity. Model economy f85p1bc0m takes the longest time to attain steady state at over 30,000 cyles. }
\end{center}

For over a hundred years, the Lorenz Curve has provided both a visual and mathematical means to compare the inequality of two different wealth distributions. The curve's construction was defined by Lorenz (1905):
\begin{quote}
\itshape
"Plot along one axis cumulated (sic) per cents. of the population from poorest to richest, and along the other the per cent. the total wealth held by these per cents. of the population."
\end{quote}
Figure 19(a) presents the Lorenz Curve for these four distributions at the end of the simulation as shown in Table 5. The model economy with highest steady state inequality is f10p1bc0m3 which has a limited number of wealth buckets, the highest turnover of agents, and the lowest mean age. Figure 19(b) presents a histogram of the individual surplus resources (wealth) for the model economies f85p1bc0m3 and f10p1bc40m3 at steady state. This figure gives a good representation of inequality by highlighting the differences between a highly unequal wealth distribution (f85p1bc0m3) and a much more equal distribution (f10p1bc40m3).

\begin{center}
	\centering
	\includegraphics[angle=-90,scale=0.8]{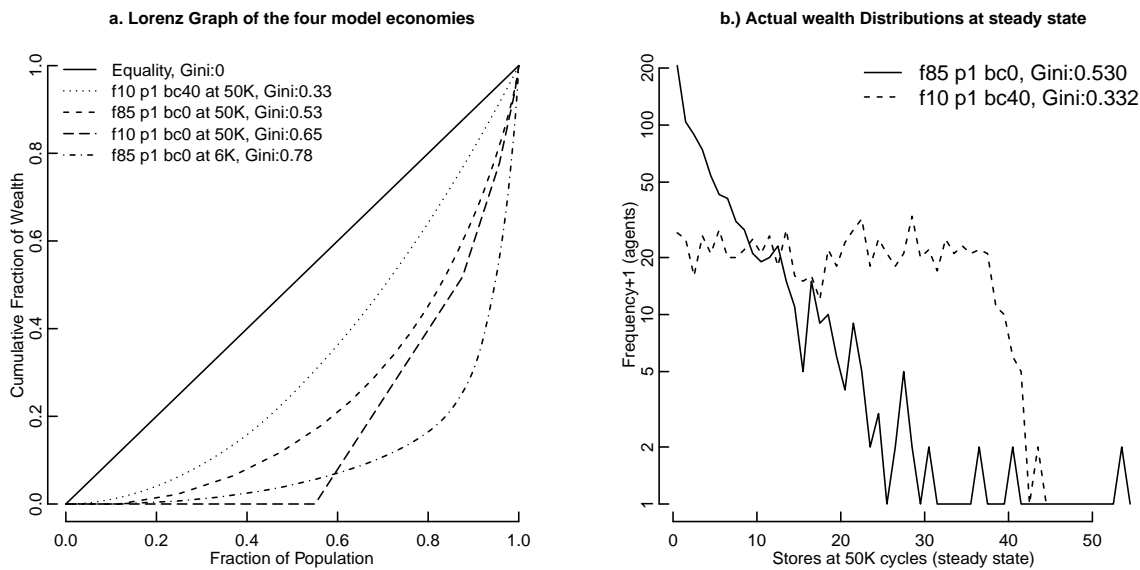}
	\captionof{figure} {a)  The Lorenz Curve of the agents' surpluses for the four model economies. b.) Histograms of the agents' surplus resources for the f85p1bc0 and f10p1bc40 surplus economies at steady state (50K cycles). The number of agents at a level of surplus resources is one less than shown on the graph (frequency +1) to enable logorithmic scaling of the vertical axis. The large difference in wealth distribution between these two simple economies is evident.}
\end{center}

Figure 20(a) provides the wealth distribution and cumulative wealth fraction for the most unequal model economy (f85p1bc0m3\_6K). This distribution, though having attained a steady population, is still far from equilibrium for mean age and wealth distribution. Its inequality is the highest, its total wealth is the highest, and its wealthiest bucket is a record 400 resources. Figure 20(b) presents the wealth distribution and cumulative wealth fraction for model economy f10p1bc0m3\_50K, which is the most unequal economy at steady state among the four model economies, has the lowest mean age for the population and, as shall be seen, has the highest death rates. The poverty of this economy is highlighted by the maximum wealth bucket of only five resources.

\begin{center}
	\centering
	\includegraphics[angle=-90,scale=0.8]{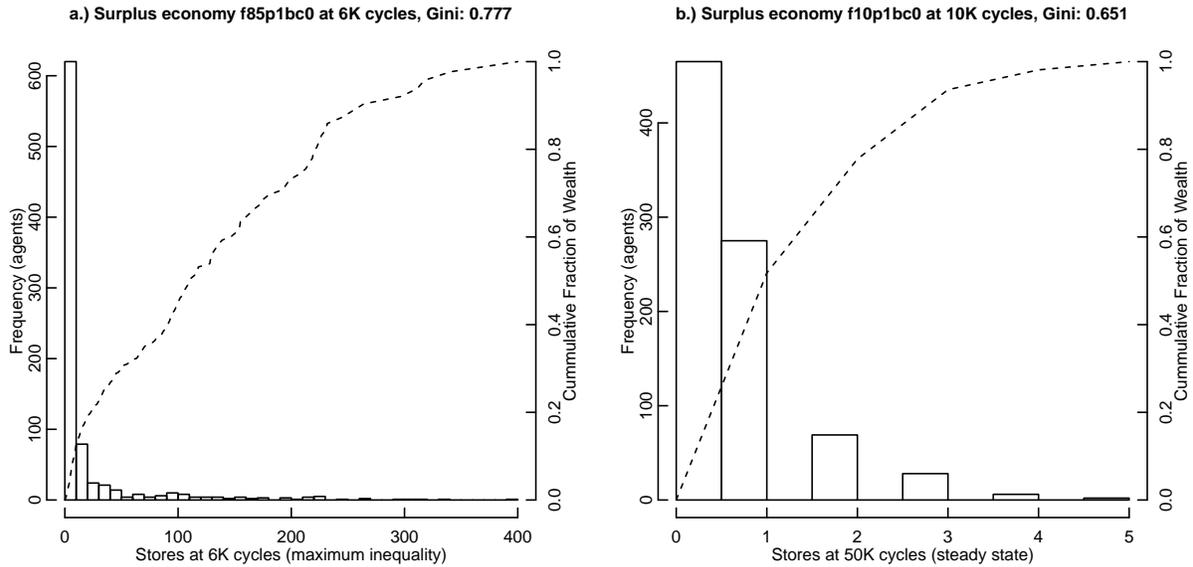}
	\captionof{figure} {a)  Histogram and cummulative fraction of agents' surplus resources for f85p1bc0 at maximum inequality (6K cycles). This distribution is far from equilibirum and represents the greatest inequality seen in the model economies.  b) Histogram and cummulative fraction of the agents' surplus resources for f10p1bc0 at steady state (50K cycles). This distribution has the smallest wealth bucket, with no agents storing more the five resources. }
\end{center}

Figure 21 presents density histograms of deaths per cycle for the three steady state model economies. These densities are taken over steady state windows of 35,000 to 50,000 cycles for f10p1bc40m3 and f85p1bc0m3, and 600 through 3,000 cycles for f10p1bc0m3. This local measure, based on the individual deaths, also gives a good approximation of the birth rate of these economies since, at steady state, the births per cycle must equal the deaths per cycle over a short time. 

\begin{center}
	\centering
	\includegraphics[angle=-90,scale=0.8]{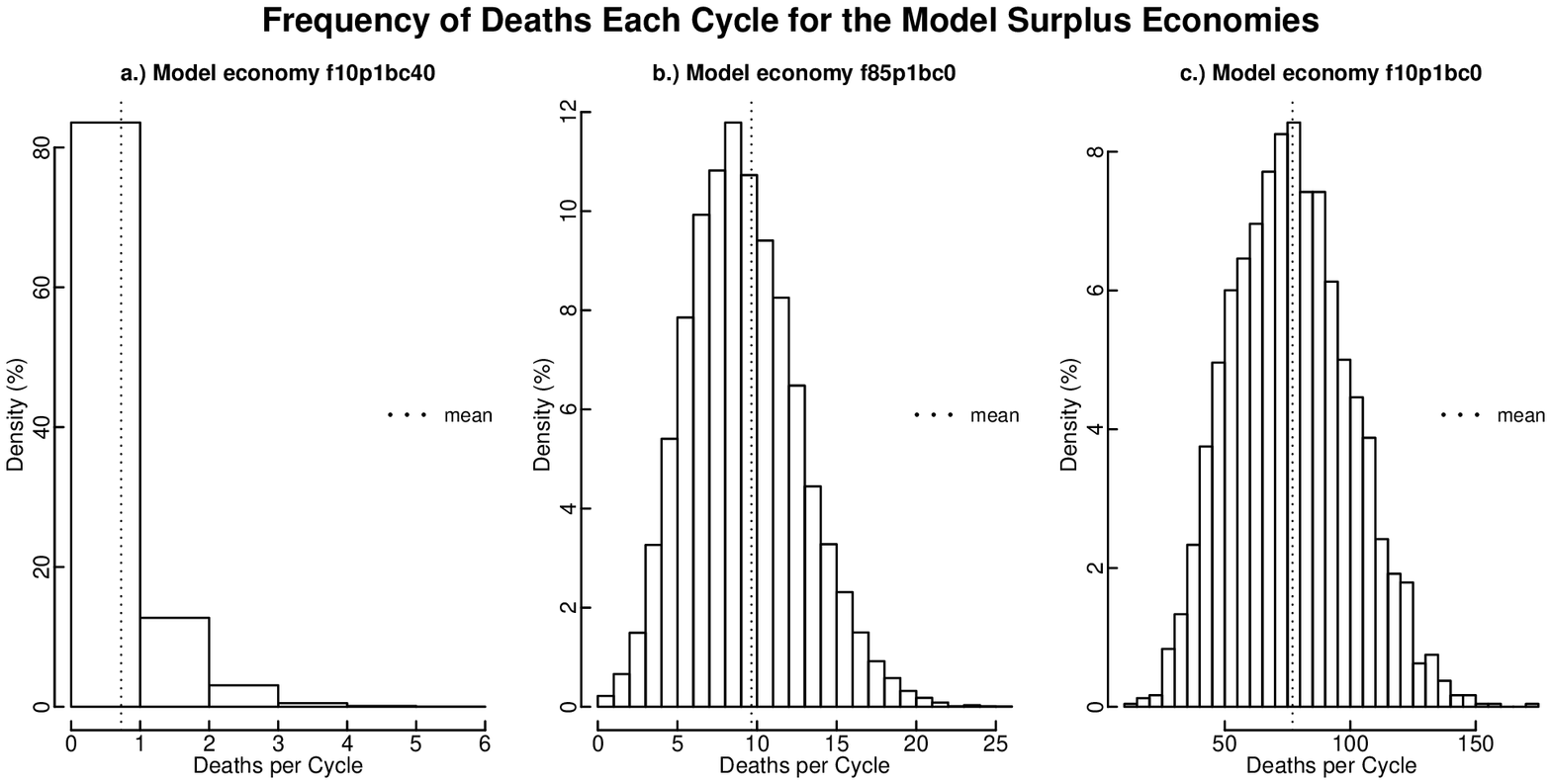}
	\captionof{figure} {The density of deaths each cycle for model surplus economies at steady state: a) f10p1bc0, b) f85p1bc0, and c) f10p1bc40. The number of deaths per cycle provide an intuitive perspective on what is happening to the population. The deaths per cycle move through two orders of magnitude from the least deaths (mean less than 1 agent per cycle) to the most (mean deaths more than 75 agents per cycle.}
\end{center}

\subsection{Non-equilibrium Inequality Dynamics}

As all these surplus model economies show, the initial agents (founders), born into a underpopulated and rich landscape, have a tremendous advantage in building up their personal wealth before the population reaches the actual carrying capacity. Figure 22 gives the initial wealth histories for the early founders of each of the three model economies. 

\begin{center}
	\centering
	\includegraphics[angle=-90,scale=0.8]{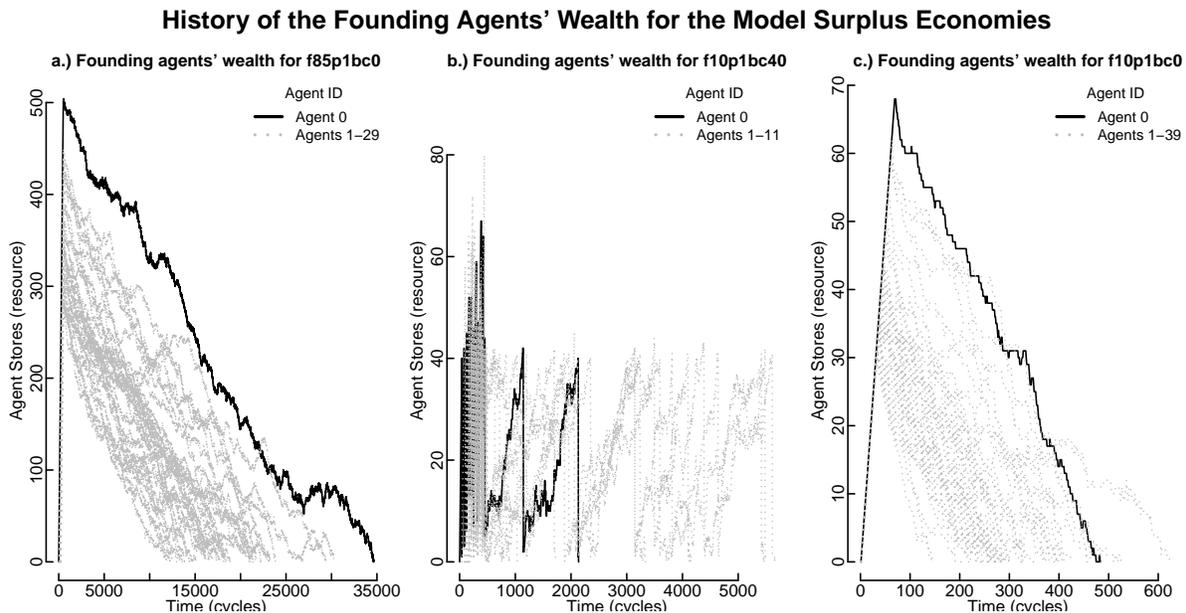}
	\captionof{figure} {a.) Wealth history for the first thirty agents of the f85p1bc0m3 economy.   b.)  Wealth history for the first twelve agents of the f10p11bc40ms economy. c) Wealth history for the first forty agents of the f10p1bc0m3 economy. The individual wealth histories of the founding agents of each model population reveal both the outsized surpluses (relative to steady state) stored by these founders as well as how long the founders' wealth persists. The wealth history of the founders of the model economy with non-zero birth cost (b) is of a much different character than the wealth histories of the founders for the other model economies (a) and (c).}
\end{center}
As Figure 18 of relaxation times previously showed, mean age and inequality of these populations do not attain steady state until these founding agents have given up their surpluses and expired, the last founder taking 35,000 cycles. The most unequal population was attained by f85p1bc0m3 at 6,000 cycles, which had a mean age of 1,327 and a total wealth of 17,132 resources [Figure 22(a)]. Even with such elite wealth and long lifespans, the peak total wealth of this economy not at equilibrium is only comparable to the steady state total wealth (16,608 cycles) and mean age (2,261) of the most equal model economy f10p1bc40m3 at steady state[Figure 22(b)]. The relationships of inequality and mean age to the growth parameters are substantially different between a given economy at steady state versus one still in transition.

\subsection{Discussion of Inequality}

Inequality emerges in populations with homogeneous populations of individuals of equal ability on landscapes providing equal opportunity. 

After a population has achieved a true steady state, the inequality, as measured by the Lorenz Curve [Figure 19(a)], has the following relationship to model parameters. Increasing birth cost [Figure 2], infertility [Figure 3(a)] and puberty [Figure 3(b)], all decrease the implied, intrinsic growth rate [Table 5]. As the intrinsic growth rate decreases, mean ages increase [Figures 14 and 15(a)], mean deaths per cycle decrease [Figure 15(b)], total wealth of the population increases [Figures 16(a)], and inequality decreases [Figure 16(b)]. These relationships suggest that large inequality at steady state is characteristic of short-lived populations driven by high death rates.

Increasing birth costs reduce inequality at a much greater rate than increasing infertility or puberty [Figure 16(b)]. The one-to-one mapping of inequality to mean age (and thereby intrinsic growth rate) is lost, suggesting an additional effect the rising birth costs have on wealth distributions. And it is perhaps counter-intuitive that these increasing birth costs do not reduce the total wealth of the population though they represent a sunk cost of wealth. These sunk costs may be compensated by the significantly lower deaths per cycle [Equation 5 and Figure 21]. Surprisingly, the high birth cost model economy (f10p1bc40m3), while the wealthiest model economy at steady state, is also the most equal model economy.

The highest levels of inequalities are found in non-equilibrium periods during and after the initial growth phase of these model economies, when a small number of agents (founders) reproduce into a rich, under-populated landscape. These founders store such significant resources before the population reaches its carry capacity that even after the carry capacity has been reached, the residual surplus resources decay quite slowly [Figure 22], preserving high inequality and preventing equilibrium for time periods greater than scale of the initial growth phase. The ability of this simple economic model to simulate these non-equilibrium dynamics provides important insights into inequality.

\section{Conclusions}

Very simple population-driven subsistence and surplus economies have been defined and then simulated with a spatial, agent-based model. This economic model is made simple by using homogeneous (equal ability) populations operating on equal opportunity (flat) landscapes. Sensitivity studies of model parameters mapped out various dynamic regimes of the population including constant level; and damped, steady, and chaotic magnitude periodic oscillations; and extinctions. Comparisons of the aggregate dynamics of these simple economies with established population-driven models from mathematical biology and ecology lend credence to the simple economic model and allowed estimation of implied, inherent rates of growth and delay periods. The extinction events of the simple economic model were examined at the local level and some were shown to occur due to spatial and stochastic characteristics whose details are only available at the individual level. These details reveal the importance of the spatial-temporal structure of populations and its stochastic nature at the individual level. By using additional measures of the distributions of agents' ages, surpluses, and deaths, the relaxation times to steady state for these economies are shown to be much greater than the time it takes for the population level to reach actual carry capacity. The driver of these long relaxation times is the outsized surplus wealth of the founders, the original colonizers of the rich, underpopulated landscape. These individual distributions of surplus resources also provide a direct measure of inequality. The sensitivity of inequality distributions to model parameters were generated and various regimes were identified. Four model economies were defined for further study and inequality distributions of these model economies were examined and compared. The dominant effect of founders on the inequality distributions before steady state is highlighted.  Even after the effects of the founders dissipated, significant inequality was shown to exist in populations of identical individuals residing in an equal opportunity (flat) landscape. The drivers of these inequalities, however, are shown to be much different in the steady state phase than during non-equilibrium transitions. The degree of inequality was shown to be a sensitive to the model parameters with slower intrinsic growth rates (with lower death rates) producing both lower inequality and greater total wealth of the population at steady state. The inclusion of birth costs made additional contributions to reduced inequality beyond its effects of reducing intrinsic growth while actually increasing the total wealth of the population even though resources were removed at every birth.

Simple economies with equal opportunity environments and equally capable individuals generate complex wealth distributions whose inequalities are dependent on the intrinsic growth rate of the population, the cost of reproduction, and whether the economy has reached equilibrium.

\section{References}
Angus SD, Hassan-Machismo B. (2015) "Anarchy" Reigns: A quantitative analysis of agent-based modelling practices in JASSS, 2001-2012. JASSS v18(4)16. DOI: 10.18564/jasss.2952.
\\ \\
Badham J, Chattoe-Brown E, Gilbert N, Chalabi Z, Kee F, Hunter RF. (2018) Developing agent-based models of complex health behaviour. Health and Place 53 170-177. https://doi.org/10.1016/j.healthplace.2018.08.022
\\ \\
Caprice A. (2011) The Ultimate Quotable Einstein. Princeton University Press.
\\ \\
Chen SH. (2012) Varieties of agents in agent-based computational economics: A historical and an interdisciplinary perspective. J. Economic Dynamics and Control 36(1) DOI:10.1016/j..2011.09.003
\\ \\
DeAngelis D, Grimm V. (2014) Individual-based models in ecology after four decades. F1000Prime v(39).doi: 10.12703/P6-39
\\ \\
Epstein JM, Axtell R. (1996) Growing Artificial Societies: Social Science from the Bottom Up. Brookings Institution Press.
\\ \\
Feldstein M. (1976) On the Theory of Tax Reform. Journal of Public Economics (6)77-104
\\ \\
Foley DK. (1999) Statistical Equilibrium in Economics: Method, interpretations, and an Example. XII Workshop on General Equilibirum at Certosa di Pontignano, Siena Italy. Corpus ID:18043050
\\ \\
Fontanari A, Taleb NN, Cirillo P. (2018) Gini Estimation Under Infinite Variance, Physica A: Statistical Mechanics and Its Applications. DOI: 10.1016/j.physa.2018.02.102
\\ \\
Gause GF. (1934) The struggle for existence. Williams and Wilkins.
\\ \\
Heppenstall A, Crooks A, Malleson N, Manley E, Ge J, Batty M. (2020) Future Developments in Geographical Agent-Based Models: Challenges and Opportunities. Geographical Analysis 0, 1–16. doi: 10.1111/gean.12267
\\ \\
Hutchinson GE, (1948) Circular causal systems in ecology. Ann New York Acad Sci. (50)221–248. DOI: 10.1111/j.1749-6632.
\\ \\
Kohler TA, Gumerman GJ. (2000) Dynamics in Human and Primate Societies:Agent-Based Modeling of Social and Spatial Processes. Santa Fe Institute.
\\ \\
Kot M. (2001) Elements of Mathematical Ecology. Cambridge University Press.
\\ \\
Lee J, Filatova T, Ligmann-Zielinska A, Hassani-Mahmooei B, Stonedahl F, Lorscheid I, Voinov A, Polhill G, Sun Z, Parker C. (2015) The complexities of Agent-Based Modeling Output Analysis. JASSS 18(4)4. DOI: 10.18564/jasss.2897
\\ \\
Lorenz MO. (1905) Methods of measuring the concentration of wealth. Publications of the American Statistical Association. V9 (New Series, No. 70) 209–219.doi:10.2307/2276207
\\ \\
Manson S, An L, Clarke KC, Heppenstall A, Koch J, Krzyzanowski B, Morgan F, O’Sullivan D, Runck BC, Shook E, Tesfatsion L. (2020) Methodological Issues of Spatial Agent-Based Models. Journal of Artificial Societies and Social Simulation 23(1) 3. Doi: 10.18564/jasss.4174 Url: http://jasss.soc.surrey.ac.uk/23/1/3.html
\\ \\
May RM.  (1974) Biological  populations  with  nonoverlapping  generations:  stable points, stable cycles, and chaos. Science 186,645–647. DOI: 10.1126/science.186.4164.645
\\ \\
Murray JD. (2002) Mathematical Biology. Springer.
\\ \\
di Porcia e Brugnera M, Fischer R, Taubert F, Huth A, Verbeeck H. (2020) Lianas in silico, ecological insights from a model of structural parasitism. Ecological Modelling 431, 109159. https://doi.org/10.1016/j.ecolmodel.2020.109159
\\ \\
Rosen HS. (1978) An Approach to the Study of Income, Utility, and Horizontal Equity. The Quarterly Journal of Economics 92(2)307-322.
\\ \\
Shoukat A, Moghadas S. (2020) Agent-Based Modelling: An Overview with Application to Disease Dynamics. arXiv:2007.04192v1 [cs.MA] 8 Jul 2020
\\ \\
Slesnick, DT. (1989) The Measurement of Horizontal Inequality. The Review of Economics and Statistics 71(3)481-490.
\\ \\
Talib NN, (2015) How to (not) estimate Gini coefficients for fat tailed variables. Tail Risk Working Paper Series, arXiv:1510.04841v1
\\ \\
Verhulst PF. (1838) notice sur la loi que la populations suit dans son accroissement. Corr Math et Phys (10)113-121.
\\ \\
Vincenot CE. (2018) How new concepts become universal scientific approaches: insights from citation network analysis of agent-based complex systems science. Proc. R. Soc. B 285: 20172360. http://dx.doi.org/10.1098/rspb.2017.2360
\\ \\
Wright EM. (1955) A  non-linear  difference-differential  equation. J für die reine und angewandte Mathematik. (194)66–87. DOI:10.1515/crll.1955.194.66
\\ \\
Yitzhaki S, Schechtmann E. (2012) The Gini Methodology: A Primer on a Statistical Methodology. Springer.

\end{document}